\documentclass[aps,prd,twocolumn,amsmath,amssymb,showpacs]{revtex4}

\usepackage{graphicx}
\usepackage{dcolumn}
\usepackage{bm}

\begin{document}

\title{Unstable circular null geodesics of static spherically symmetric black holes, \\
Regge poles and quasinormal frequencies}

\author{Yves D\'ecanini}
\email{decanini@univ-corse.fr}
\author{Antoine Folacci}
\email{folacci@univ-corse.fr}
\author{Bernard Raffaelli}
\email{raffaelli@univ-corse.fr}
\affiliation{
UMR CNRS 6134 SPE, Equipe Physique Th\'eorique, \\
Universit\'e de Corse, Facult\'e des Sciences, BP 52, 20250 Corte,
France}

\date{\today}

\begin{abstract}

We consider a wide class of static spherically symmetric black holes
of arbitrary dimension with a photon sphere (a hypersurface on which
a massless particle can orbit the black hole on unstable circular
null geodesics). This class includes various spacetimes of physical
interest such as Schwarzschild, Schwarzschild-Tangherlini and
Reissner-Nordstr\"om black holes, the canonical acoustic black hole
or the Schwarzschild-de Sitter black hole. For this class of black
holes, we provide general analytical expressions for the Regge poles
of the $S$-matrix associated with a massless scalar field theory.
This is achieved by using third-order WKB approximations to solve
the associated radial wave equation. These results permit us to
obtain analytically the nonlinear dispersion relation and the
damping of the ``surface waves" lying close to the photon sphere as
well as, from Bohr-Sommerfeld--type resonance conditions, formulas
beyond the leading-order terms for the complex frequencies
corresponding to the weakly damped quasinormal modes.

\end{abstract}

\pacs{04.70.-s, 04.50.Gh}

\maketitle

\section{Introduction}

Quasinormal modes (QNMs) of black holes (BHs) have been studied for
nearly 40 years due to their importance in the context of
gravitational wave astronomy. In the last decade, there has been
moreover an increase of activity in BH QNM studies motivated by
potential applications in analog models of gravity, quantum gravity,
string theory and related topics (TeV-scale gravity, AdS/CFT
correspondence, alternative theories of gravity, BH area
quantization, phase transitions in BH systems,...). For excellent
reviews on the status of QNMs prior to 1999 and on their relevance
to gravitational wave astronomy, we refer to the articles by
Kokkotas and Schmidt \cite{KokkotasSchmidt99} and by Nollert
\cite{Nollert99}. For a more recent review on BH QNMs, we refer to
the article by Berti, Cardoso and Starinets
\cite{BertiCardosoStarinets2009}: it updates the two previously
cited articles and it also presents the aspects of QNM physics
linked to gauge-gravity duality; it includes, furthermore, an
interesting historical introduction on the subject as well as a
useful impressive bibliography on all the aspects of BH physics
linked to QNMs.

Immediately after the publication of one of the first papers on QNMs
by Press \cite{Press1971} where he identified the gravitational
ringing of the Schwarzschild BH as due to its ``free oscillations",
Goebel suggested a physically intuitive interpretation of the
associated QNMs \cite{Goebel}: they could be interpreted in terms of
gravitational waves in spiral orbits close to the unstable circular
photon/graviton orbit at $r=3M$ which decay by radiating away energy
(here $M$ denotes the mass of the BH). This appealing interpretation
has been developed by other authors for various field theories
defined on BH backgrounds using the eikonal approximation, i.e., in
a framework based on geodesics and bundle of geometrical rays (see
Refs.~\cite{FerrariMashhoon84,Mashhoon85,Stewart89,
AnderssonOnozawa96,ZerbiniVanzo2004,CardosoMirandaBertietal2009,DolanOttewill_2009,Hod2009}
as well as Ref.~\cite{BarretoZworski1997} for a more mathematical
approach). It has permitted them to obtain analytical approximations
for the leading-order terms of the characteristic complex
frequencies of various BH spectra from an interpretation in terms of
massless particles ``trapped" near unstable circular null geodesics
(see, more particularly, Ref.~\cite{CardosoMirandaBertietal2009}
where the relation with the Lyapunov exponent corresponding to
geodesic motion is clearly emphasized).

A potentially much richer implementation of the Goebel
interpretation of BH QNMs which is not limited to purely geometrical
considerations but based on wave/field theory and which goes beyond
the leading oder terms has also been formulated
\cite{DecaniniFJ_cam_bh,DecaniniFolacci2009,DecaniniFolacci2010a}
(see also Ref.~\cite{DolanOttewill_2009}). It uses complex angular
momentum (CAM) techniques (or, in other words, the Regge pole
machinery) which play a central role in scattering theory. Since, as
noted by Chandrasekhar and coworkers \cite{Chandrasekhar_1983} (see
also Ref.~\cite{Ferrari1992}), BH perturbation theory can be
formulated as a resonant scattering problem, CAM techniques arise
naturally in BH physics. For reviews of the CAM method, we refer to
the monographs of Newton \cite{New82}, Nussenzveig \cite{Nus92} and
Collins \cite{Collins77} as well as to references therein for
various applications in quantum mechanics, nuclear physics, high
energy physics, electromagnetism and seismology.

Some years ago, the CAM method was used in gravitational wave
physics by Chandrasekar and Ferrari \cite{Chandra} to express the
flow of energy due to nonradial oscillations of relativistic stars
and by Andersson and Thylwe to describe scattering from the
Schwarzschild BH \cite{Andersson1} as well as to interpret the
Schwarzschild BH glory \cite{Andersson2}. In this context, Andersson
established, for the Schwarzschild BH of mass $M$, the existence of
a family of ``surface waves" (each one associated with a Regge pole
of the $S$-matrix) orbiting close to the unstable photon orbit at
$r=3M$ (see also Ref.~\cite{DecaniniFJ_cam_bh} for a more rigorous
approach). Recently, from these ``surface waves", we have been able
to theoretically and numerically construct the spectrum of the
weakly damped complex frequencies of the Schwarzschild BH QNMs
\cite{DecaniniFJ_cam_bh,DecaniniFolacci2010a} and to interpret them
as Breit-Wigner resonances. This has been achieved by obtaining
analytically the nonlinear dispersion relation as well as the
damping of the ``surface waves" propagating close to the photon
sphere. Let us also note two related papers concerning analytical or
numerical determinations of the Regge poles of the Schwarzschild BH
\cite{GlampedakisAndersson2003,DolanOttewill_2009} and that, this
last year, Regge poles have also been used to understand the
resonant aspects of the BTZ BH \cite{DecaniniFolacci2009} as well as
to analyze some aspects of self-force calculations
\cite{Casals_et_al_2009}.

In the present paper, we extend the analysis developed for the
Schwarzschild BH to more general BHs and we establish, from Regge
pole considerations, a precise connection between the existence of a
photon sphere and the properties of the ``surface waves" propagating
close to it. More precisely, we consider a wide class of static
spherically symmetric BHs of arbitrary dimension with a photon
sphere, i.e., a hypersurface on which a massless particle can orbit
the BH on unstable circular null geodesics. For more rigorous
definitions of the photon sphere concept in static spherically
symmetric spacetime, we refer to the article by Claudel, Virbhadra
and Ellis \cite{ClaudelVirbhadraEllis2001}. This class of BHs
includes various spacetimes of physical interest such as
Schwarzschild, Schwarzschild-Tangherlini and Reissner-Nordstr\"om
BHs, the canonical acoustic BH or the Schwarzschild-de Sitter BH.
For this class of BHs, we provide general analytical expressions
beyond the leading-order terms for the Regge poles of the $S$-matrix
associated with a massless scalar field theory. These results permit
us to obtain analytically the nonlinear dispersion relation and the
damping of the ``surface waves" lying close to the photon sphere as
well as, from Bohr-Sommerfeld--type resonance conditions, the
complex frequencies corresponding to the weakly damped QNMs.

Our paper is organized as follows. In Sec.~II, we display our
general working assumptions and we justify them physically. We then
explain how to construct the $S$-matrix permitting us to analyze the
resonant aspects of a scalar field theory defined on an
asymptotically flat static spherically symmetric BH of arbitrary
dimension with a photon sphere and we finally define its Regge poles
as well as its complex quasinormal frequencies. In Sec.~III, we
provide a general analytical expression for the Regge poles. This is
achieved by using and extending the WKB approach developed in the
context of the determination of the QNMs by Schutz and Will
\cite{SchutzWill} and by Will and Iyer \cite{Iyer1,Iyer2} (see also
Ref.~\cite{BenderOrszag1978} for general aspects of WKB theory and
for particular aspects connected with eigenvalue problems). Our
result permits us to describe the Regge trajectories of a general
asymptotically flat static spherically symmetric BH of arbitrary
dimension with a photon sphere and to obtain, from semiclassical
formulas, analytical expressions for the QNM complex frequencies.
Our WKB analysis permits us moreover to show that (i) the dispersion
relation of the $n$th ``surface wave" is nonlinear and depends on
the index $n$ and that (ii) the damping of the $n$th ``surface wave"
is frequency dependent. In Sec.~IV, we apply the general theory
developed in Sec.~III to particular BHs (Schwarzschild,
Schwarzschild-Tangherlini, Reissner-Nordstr\"om and canonical
acoustic BHs). In a brief conclusion, we consider some consequences
of our work as well as possible extensions. In Appendix A, we
establish the semiclassical connection between the Regge poles of a
static spherically symmetric BH of arbitrary dimension with a photon
sphere and the complex frequencies of its weakly damped QNMs. In
Appendix B, we consider the particular case of the Schwarzschild-de
Sitter BH. Indeed, even if such a gravitational background is not
asymptotically flat, the formalism developed in Secs.~II and III
naturally applies to it.

In this paper, we shall use units with $\hbar=c=G=1$.

\section{Quasinormal frequencies and Regge poles of static
spherically symmetric black holes: General theory}

We consider a static spherically symmetric spacetime of arbitrary dimension $d \ge 4$
with metric
\begin{equation} \label{metric_BH}
ds^2=-f(r)dt^2+\frac{dr^2}{f(r)}+r^2d\sigma_{d-2}^{2}.
\end{equation}
Here $d\sigma_{d-2}^{2}$ denotes the line element on the unit sphere
$S^{d-2}$. On $S^{d-2}$, we introduce the usual angular coordinates
$\theta_i \in [0,\pi]$ with $i=1, \dots, d-3$ and $\varphi \in
[0,2\pi]$. We have
\begin{equation} \label{metric_Sphere}
d\sigma_{d-2}^{2}={d\theta_1}^2 + \sum_{k=2}^{d-3} \left(\prod_{i=1}^{k-1}
\sin^2\theta_i\right)d\theta_k^2 + \left(\prod_{i=1}^{d-3}
\sin^2\theta_i\right)d\varphi^2.
\end{equation}
Of course, a metric such as (\ref{metric_BH}) does not describe the
most general static spherically symmetric spacetime but it will
permit us to consider a wide class of BHs of physical interest.

In Eq.~(\ref{metric_BH}), we shall furthermore assume that $f(r)$
is a function of the usual radial coordinate
$r$ with the following properties:

\begin{itemize}
\item (i) There exists an interval $I=]r_h,+\infty[ \subset \bf{R}$
with $r_h>0$ such as $f(r)>0$ for $r \in I$.
\item (ii) $r_h$ is a simple root of $f(r)$, i.e.,
\begin{equation} \label{Assup_fr_1a}
f(r_h)=0 \qquad \text{and} \qquad f'(r_h)\neq 0,
\end{equation}
and $f(r)$ moreover satisfies
\begin{equation}
\underset{r \to +\infty}{\lim}f(r)=1.
\end{equation}
\item (iii) There exists a value $r_c \in I$ for which
\begin{equation} \label{Assup_fr_2a}
f'(r_c) - \frac{2}{r_c}f(r_c)=0
\end{equation}
and
\begin{equation} \label{Assup_fr_2b}
f''(r_c)-\frac{2}{r_c^2}f(r_c) <0.
\end{equation}
\end{itemize}

We shall now briefly discuss assumptions (i)-(iii) previously
introduced. Assumptions (i) and (ii) indicate that the spacetime
considered is an asymptotically flat BH with an event horizon at
$r_h$, its exterior corresponding to $r \in I$. Assumption (iii)
implies the existence of a photon sphere which is the support of
unstable circular null geodesics (see below for more details). It
should be noted that, as a consequence of (i) and (ii), the tortoise
coordinate $r_\ast=r_\ast(r)$ defined for $r \in I$ by the relation
$dr_\ast/dr=1 / f(r)$ and the condition $r_\ast(r_c)=0$ provides a
bijection $r_\ast=r_\ast(r)$ from $I$ to $]-\infty,+\infty[$.

Let us consider a free-falling massless particle orbiting the BH.
Without loss of generality, we can consider that its motion lies on
the equatorial hyperplane defined by $\theta_i=\pi/2$ for $i=1,
\dots, d-3$. Because it moves along a null geodesic, we have [cf.
Eqs.~(\ref{metric_BH}) and (\ref{metric_Sphere})]
\begin{equation}\label{nullgeod}
-f(r)\left(\frac{dt}{d\alpha}\right)^2
+\frac{1}{f(r)}\left(\frac{dr}{d\alpha}\right)^2+r^2\left(\frac{d\varphi}{d\alpha}\right)^2=0
\end{equation}
where $\alpha$ is an affine parameter and, of course, there exist
two integrals of motion respectively associated with the Killing
vectors $\partial / \partial t$ and $\partial / \partial \varphi$
and given by
\begin{subequations}
\begin{eqnarray}
&&f(r)\left(\frac{dt}{d\alpha}\right)=E,\label{motioncst1}\\
&&r^2\left(\frac{d\varphi}{d\alpha}\right)=L.\label{motioncst2}
\end{eqnarray}
\end{subequations}
Here $E$ and $L$ denote respectively the energy and the angular momentum of the massless particle.
Inserting Eqs.~(\ref{motioncst1}) and (\ref{motioncst2}) into (\ref{nullgeod}), we obtain
\begin{equation}\label{eqmotion}
\left(\frac{dr}{d\alpha}\right)^2+V_{\mathrm{eff}}(r)=E^2
\end{equation}
where
\begin{equation}\label{Veff}
V_{\mathrm{eff}}(r)=\frac{L^2}{r^2}f(r).
\end{equation}
From these last two equations and from assumption (iii), one can
easily remark that the massless particle can orbit the BH on an
unstable circular geodesic defined by $r=r_c$. Indeed, we have in
particular
\begin{subequations}
\begin{equation}\label{Veff_d1}
\left. \frac{d}{dr}V_{\mathrm{eff}}(r)\right|_{r=r_c}=0
\end{equation}
and
\begin{equation}\label{Veff_d2}
\left.
\frac{d^2}{dr^2}V_{\mathrm{eff}}(r)\right|_{r=r_c}=\frac{L^2}{r_c^2}\left(f''(r_c)-\frac{2}{r_c^2}f(r_c)
\right) <0.
\end{equation}
\end{subequations}
On this orbit, the massless particle takes the time
\begin{equation}\label{Torbit}
T=\frac{2\pi r_c}{\sqrt{f(r_c)}}
\end{equation}
to circle the BH. This result can be obtained by integrating Eq.~(\ref{nullgeod}).

The wave equation for a massless scalar field propagating on a
general gravitational background is given by
\begin{equation}\label{WaveEq}
\Box \Phi=g^{\mu\nu}\nabla_{\mu}\nabla_{\nu}\Phi=
\frac{1}{\sqrt{-g}}\partial_{\mu}\left(\sqrt{-g}g^{\mu\nu}\partial_{\mu}\Phi\right)=0.
\end{equation}
If the spacetime metric is given by (\ref{metric_BH}), after
separation of variables and the introduction of the radial partial
wave functions $\Phi_\ell(r)$ with $\ell=0,1,2, \dots$, this wave
equation reduces to the Regge-Wheeler equation
\begin{equation}\label{RW}
\frac{d^2 \Phi_\ell}{dr_*^2} + \left[ \omega^2 - V_\ell(r)\right]
\Phi_\ell=0.
\end{equation}
[Here we have assumed a harmonic time dependence $\exp(-i\omega t)$
for the massless scalar field.] In Eq.~(\ref{RW}), $V_\ell(r)$ is
the Regge-Wheeler potential given by
\begin{eqnarray}\label{EffectivePot}
V_\ell(r)=f(r) \left[ \frac{\ell(\ell+d-3)}{r^2} \right. &+& \left.
\frac{(d-2)(d-4)}{4r^2}f(r)\right. \nonumber \\
&+& \left. \left(\frac{d-2}{2r}\right)f'(r) \right].
\end{eqnarray}
It should be noted that
\begin{itemize}
\item $\lim_{r \to r_h} V_\ell(r)=0$ and $\lim_{r \to +\infty} V_\ell(r)=0$
and therefore the solutions of the radial equation (\ref{RW}) have a behavior
in $\exp(\pm i \omega r_{\ast})$ at the horizon and at infinity.
\item For $\ell \gg 1$, $V_{\ell}(r)$ has a local maximum at $r=r_c$ because,
in this limit, $V_{\ell}(r)$ and $V_{\mathrm{eff}}(r)$ are similar.
\item For any finite value of $\ell$, the local maximum of $V_{\ell}(r)$ is close to $r=r_c$.
\end{itemize}

For a given angular momentum index $\ell$, the $S$-matrix element
$S_\ell (\omega)$ is defined by seeking the solution of the
Regge-Wheeler equation (\ref{RW}) which has a purely ingoing
behavior at the event horizon $r=r_h$, i.e., which satisfies
\begin{equation}\label{bc1}
\Phi_\ell (r) \underset{r_* \to -\infty}{\sim}
T_\ell(\omega)e^{-i\omega r_* }
\end{equation}
and which, at spatial infinity $r \to +\infty$, presents an
asymptotic behavior of the form
\begin{eqnarray}\label{bc2}
&& \Phi_\ell(r) \underset{r_* \to +\infty}{\sim}
e^{-i\omega r_* + i\left(\ell+\frac{d-3}{2}\right)\frac{\pi}{2}-i\frac{\pi}{4}} \nonumber \\
&& \qquad \qquad\qquad  -S_\ell(\omega)e^{+i\omega r_* -
i\left(\ell+\frac{d-3}{2}\right)\frac{\pi}{2}+i\frac{\pi}{4}}.
\end{eqnarray}
We recall that the $S$-matrix permits us to analyze the resonant
aspects of the considered BH as well as to construct the
form factor describing the scattering of a monochromatic scalar wave (see Appendix A).

To describe semiclassically resonance phenomena, the dual structure
of the $S$-matrix plays a crucial role. Indeed, the $S$-matrix is a
function of both the frequency $\omega$ and the angular momentum
index $\ell$. It can be analytically extended into the complex
$\omega$-plane as well as into the complex $\lambda$-plane (CAM
plane) with $\lambda=\ell+(d-3)/2$. From now on, we shall denote by
$S_{\lambda -(d-3)/2}(\omega)$ this double analytical extension. For
$\ell=\lambda -(d-3)/2 \in \bf{N}$, the simple poles lying in the
fourth quadrant of the complex $\omega$-plane [let us recall that
they are also simple poles of $T_{\lambda -(d-3)/2}(\omega)$] are
the complex frequencies of the QNMs. These modes are therefore
solutions of the radial wave equation (\ref{RW}) which are purely
outgoing at infinity and purely ingoing at the horizon. We shall
denote by $\omega_{\ell n}=\omega^{(o)}_{\ell n}- i\Gamma_{\ell
n}/2$ where $n\in {\bf N}^\ast$ the quasinormal frequencies. We
recall that $\omega^{(o)}_{\ell n}>0$ and $\Gamma_{\ell n}>0$
represent respectively the frequency of the oscillation and the
damping corresponding to the associated QNM. We assume that, in the
immediate neighborhood of $\omega_{\ell n}$, $S_\ell(\omega)$ has
the Breit-Wigner form, i.e.,
\begin{equation}\label{BW}
S_\ell(\omega ) \propto \frac{\Gamma _{\ell n}/2}{\omega
-\omega^{(o)}_{\ell n}+i\Gamma _{\ell n}/2}.
\end{equation}
For a given value $\omega >0$ of the frequency, the simple poles lying
in the first quadrant of the complex $\lambda$-plane are the so-called Regge poles.
It should be noted that they are also poles of $T_{\lambda -(d-3)/2}(\omega)$ and therefore
the associated modes (Regge modes) are purely outgoing at infinity and purely ingoing at
the horizon. We shall denote the Regge poles by $\lambda_n(\omega)$, the index $n=1,2,\dots$
permitting us to distinguish each pole.

The structure of the $S$-matrix in the complex $\lambda$-plane
allows us, by using integration contour deformations, Cauchy's
theorem and asymptotic analysis, to provide a semiclassical
description of scattering (see Appendix A for more precisions). The
curves traced out in the CAM plane by the Regge poles as a function
of the frequency $\omega$ are the so-called Regge trajectories. They
permit us to interpret Regge poles in terms of ``surface waves" (see
Appendix A): $\mathrm{Re} \, \lambda_n (\omega)$ provides the
dispersion relation for the $n$th ``surface wave" while $\mathrm{Im}
\, \lambda_n (\omega)$ corresponds to its damping. Furthermore, from
the Regge trajectories, we can semiclassically construct the
resonance spectrum [see, in Appendix A, formulas
(\ref{sc1_Appendix}), (\ref{sc2_Appendix}) and
(\ref{sc4_Appendix})]. The semiclassical formula (a
Bohr-Sommerfeld--type quantization condition)
\begin{equation}\label{sc1}
\mathrm{Re}  \, \lambda_n \left(\omega^{(0)}_{\ell n} \right)= \ell
+ \frac{d-3}{2},  \qquad \ell \in {\bf N}
\end{equation}
provides the location of the excitation frequencies $\omega^{(0)}_{\ell n}$ of
the resonances generated by $n$th
``surface wave", while a second semiclassical formula gives the widths of these resonances
\begin{equation}\label{sc2} \frac{\Gamma _{\ell n}}{2}= \left.  \frac{\mathrm{Im} \,
\lambda_n (\omega )[d/d\omega \, \mathrm{Re}\, \lambda_n (\omega )
]}{[d/d\omega \, \mathrm{Re} \, \lambda_n (\omega ) ]^2 + [d/d\omega
\, \mathrm{Im} \, \lambda_n (\omega ) ]^2 } \right|_{\omega
=\omega^{(0)}_{\ell n}}.
\end{equation}
It should be moreover noted that this formula reduces, in the
frequency range where the condition $| d/d\omega \,\mathrm{Re} \,
\lambda_n (\omega )  | \gg |d/d\omega \, \mathrm{Im}\, \lambda_n
(\omega )  |$ is satisfied, to
\begin{equation}\label{sc4}
\frac{\Gamma _{\ell n}}{2}= \left.  \frac{\mathrm{Im} \, \lambda_n
(\omega )}{d/d\omega \,   \mathrm{Re} \, \lambda_n (\omega )  }
\right|_{\omega =\omega^{(0)}_{\ell n}}.
\end{equation}

\section{WKB approximations for the Regge poles and
semiclassical expressions of the complex quasinormal frequencies}

In general, it is not possible to solve exactly the Regge-Wheeler
equation (\ref{RW}) and therefore we can obtain only analytical
approximations for the Regge poles and for the complex quasinormal
frequencies. For example, the WKB approach developed in the general
context of eigenvalue problems (for more details see
Ref.~\cite{BenderOrszag1978}) has been adapted for the determination
of the Schwarzschild BH QNMs by Schutz and Will \cite{SchutzWill}
and by Will and Iyer \cite{Iyer1,Iyer2} and, in
Ref.~\cite{DecaniniFolacci2010a}, for the determination of the
Schwarzschild BH Regge poles. It can be extended to the more general
case considered in this paper. By using third-order WKB
approximations \cite{Iyer1,Iyer2} for the Regge modes of
Eq.~(\ref{RW}), we find that the Regge poles $\lambda$ are the
complex solutions of the equation
\begin{eqnarray}\label{WKB_RP_1}
& &\omega^2=\left[V_0(\lambda) + {[-2V_0^{(2)}(\lambda)]}^{1/2} \,
{\overline \Lambda}(\lambda,n) \right] \nonumber \\ && \qquad - i \,
\alpha (n) \, {[-2V_0^{(2)}(\lambda)]}^{1/2}[1+ {\overline
\Omega}(\lambda ,n)]
\end{eqnarray}
with $\omega >0$ and $n=1,2,3, \dots $. Here
\begin{subequations}\label{WKB_RP_2}
\begin{eqnarray}\label{WKB_RP_2a}
& & {\overline
\Lambda}(\lambda,n)=\frac{1}{{[-2V_0^{(2)}(\lambda)]}^{1/2}}
\left[\frac{1}{8}
\frac{V_0^{(4)}(\lambda)}{V_0^{(2)}(\lambda)}\left( \frac{1}{4}+
\alpha(n)^2 \right) \right. \nonumber \\
& & \qquad \left.  -\frac{1}{288}
\left(\frac{V_0^{(3)}(\lambda)}{V_0^{(2)}(\lambda)}\right)^2\left(
7+ 60 \, \alpha(n)^2 \right) \right]
\end{eqnarray}
and
\begin{eqnarray}\label{WKB_RP_2b}
& &{\overline \Omega}(\lambda,n)=\frac{1}{[-2V_0^{(2)}(\lambda)]} \times  \nonumber \\
&& \qquad \left[\frac{5}{6912}
\left(\frac{V_0^{(3)}(\lambda)}{V_0^{(2)}(\lambda)}\right)^4\left(
77+ 188 \, \alpha(n)^2 \right) \right. \nonumber \\
&& \qquad \left. -\frac{1}{384}
\left(\frac{{[V_0^{(3)}(\lambda)]}^2V_0^{(4)}(\lambda)}{{[V_0^{(2)}(\lambda)]}^3}\right)
\left(
51+ 100 \, \alpha(n)^2 \right)\right. \nonumber \\
&& \qquad \left. +  \frac{1}{2304}
\left(\frac{V_0^{(4)}(\lambda)}{V_0^{(2)}(\lambda)}\right)^2 \left(
67+ 68 \, \alpha(n)^2 \right) \right. \nonumber \\
&& \qquad \left. +\frac{1}{288} \left(\frac{ V_0^{(3)}(\lambda)
V_0^{(5)}(\lambda)}{{[V_0^{(2)}(\lambda)]}^2}\right) \left( 19+ 28
\, \alpha(n)^2 \right)  \right. \nonumber \\
&& \qquad \left. -\frac{1}{288}
\left(\frac{V_0^{(6)}(\lambda)}{V_0^{(2)}(\lambda)}\right)\left( 5 +
4 \, \alpha(n)^2 \right) \right].
\end{eqnarray}
\end{subequations}
In Eqs.~(\ref{WKB_RP_1}) and (\ref{WKB_RP_2}), we have
introduced the notations
\begin{equation}\label{WKB_RP_3a}
\alpha(n)=n-1/2
\end{equation}
and, for $p \in {\bf N}$,
\begin{equation}\label{WKB_RP_3b}
V_0^{(p)} (\lambda) = \left.  \frac{d^p}{{dr_*}^p }
V_{\lambda-(d-3)/2}(r_*)\right|_{r_*={(r_*)}_0}
\end{equation}
with ${(r_*)}_0$ which denotes the maximum of the function
$V_{\lambda-(d-3)/2}(r_*)$.

In order to solve Eq.~(\ref{WKB_RP_1}), we need to express the asymptotic
expansions for $|\lambda| \to +\infty$ of $V_0(\lambda)$ and
$\left[-2V_{0}^{(2)}(\lambda)\right]^{1/2}$ and of the various ratios
appearing in Eqs.~(\ref{WKB_RP_2a}) and (\ref{WKB_RP_2b}). In order to simplify
the results we have introduced the notations
\begin{equation}
f_c^{(p)}=f^{(p)}(r_c)
\end{equation}
and
\begin{equation}\label{Eta}
\eta_c=\frac{1}{2}\sqrt{4f_{c}-2r_{c}^{2}f_{c}^{(2)}}.
\end{equation}
It is worth noting that the $\eta_c$ parameter is directly linked to
the second derivative (\ref{Veff_d2}) of the effective potential
(\ref{Veff}) taken at $r=r_c$. As a consequence, it represents a
kind of measure of the instability of the circular orbits lying on
the photon sphere. In fact, it can be expressed in terms of the
Lyapunov exponent $\Lambda_c$ corresponding to these orbits
introduced in Ref.~\cite{CardosoMirandaBertietal2009} and which is
the inverse of the instability time scale associated with them: we
have
\begin{equation}\label{Eta_c_Lyapunov}
\eta_c= \frac{r_c}{\sqrt{f_c}}|\Lambda_c|.
\end{equation}
We will say no more about this connection because, as already
mentioned in Sec.~I, we intend to go beyond purely geometrical
considerations in our analysis of the resonant behavior of BHs.

After a tedious calculation, we obtain
\begin{widetext}
\begin{subequations}
\begin{equation}\label{DevEffPot}
V_{0}(\lambda)=\lambda^{2}\frac{f_{c}}{r_{c}^{2}}+\frac{f_{c}}{4r_{c}^{2}}
\left[d(d-2)f_{c}-(d-3)^{2}\right]+\underset{|\lambda| \to
+\infty}{\cal O}\left( \frac{1}{\lambda^2}\right)
\end{equation}
and
\begin{eqnarray}\label{DevDerSecPot}
\left[-2V_{0}^{(2)}(\lambda)\right]^{1/2}&=& \frac{2\eta_c
f_{c}}{r_{c}^{2}} \lambda -\frac{f_{c}}{8 \eta_c^{3}r_{c}^{2}}\left[
2f_c^2[(d-3)^{2}+(d-2)(d-4)f_c] \phantom{\left(f_{c}^{(2)}\right)^{2}}\right. \nonumber
\\ && \left. +r_{c}^{2}f_{c}f_{c}^{(2)}
[(d-2)(d+8)f_c-2(d-3)^2]+d(d-2)r_{c}^{3}f_{c}^{2}f_{c}^{(3)} \phantom{\left(f_{c}^{(2)}\right)^{2}}\right. \nonumber
\\ && \left. +r_{c}^{4}
\left(f_{c}^{(2)}\right)^{2}[(1/2)(d-3)^2-(d^2-4)f_c]\right]\frac{1}{\lambda}
+\underset{|\lambda| \to +\infty}{\cal O}\left(
\frac{1}{\lambda^3}\right)
\end{eqnarray}
\end{subequations}
as well as
\begin{subequations}
\begin{eqnarray}\label{RapportWKB_1}
\frac{V_{0}^{(4)}(\lambda)}{V_{0}^{(2)}(\lambda)}&=&-\frac{f_{c}}{2\eta_c^{2}r_{c}^{2}}
\left[16f_{c}^{2}-16r_{c}^{2}f_c
f_{c}^{(2)}+4r_{c}^{3}f_{c}f_{c}^{(3)}
\phantom{\left(f_{c}^{(2)}\right)^{2}}\right. \nonumber
\\ && \left.+r_{c}^{4}\left(4\left(f_{c}^{(2)}\right)^{2}+f_{c}f_{c}^{(4)}\right)\right]+\underset{|\lambda| \to +\infty}{\cal O}\left(
\frac{1}{\lambda^2}\right),
\end{eqnarray}
\begin{eqnarray}\label{RapportWKB_2}
\left(\frac{V_{0}^{(3)}(\lambda)}{V_{0}^{(2)}(\lambda)}\right)^{2}=
\frac{r_{c}^{4}f_{c}^{2}\left(f_{c}^{(3)}\right)^{2}}{4\eta_c^{4}}
+\underset{|\lambda| \to +\infty}{\cal O}\left( \frac{1}{\lambda^2}\right),
\end{eqnarray}
\begin{eqnarray}\label{RapportWKB_4}
\frac{{[V_{0}^{(3)}(\lambda)]}^{2}V_{0}^{(4)}(\lambda)}{{\left[V_{0}^{(2)}(\lambda)\right]}^{3}}
&=&-\frac{r_{c}^{2}f_{c}^{3}\left(f_{c}^{(3)}\right)^{2}}{8\eta_c^{6}}\left[16f_{c}^{2}
-16r_{c}^{2}f_cf_{c}^{(2)}+4r_{c}^{3}f_{c}f_{c}^{(3)}
\phantom{\left(f_{c}^{(2)}\right)^{2}}\right. \nonumber
\\ && \left.+r_{c}^{4}\left(4\left(f_{c}^{(2)}\right)^{2}
+f_{c}f_{c}^{(4)}\right)\right]+\underset{|\lambda| \to
+\infty}{\cal O}\left( \frac{1}{\lambda^2}\right),
\end{eqnarray}
\begin{eqnarray}\label{RapportWKB_6}
\frac{V_{0}^{(3)}(\lambda)V_{0}^{(5)}(\lambda)}{{\left[V_{0}^{(2)}(\lambda)\right]}^{2}}
&=&\frac{r_{c}^{2}f_{c}^{3}f_{c}^{(3)}}{4\eta_c^{4}}\left[-10f_{c}f_{c}^{(3)}+10r_{c}f_{c}f_{c}^{(4)}\right.
\nonumber
\\ && \left.+r_{c}^{2}\left(15f_{c}^{(2)}f_{c}^{(3)}+f_{c}f_{c}^{(5)}\right)\right]
+\underset{|\lambda| \to +\infty}{\cal O}\left( \frac{1}{\lambda^2}\right),
\end{eqnarray}
\begin{eqnarray}\label{RapportWKB_7}
& & \frac{V_{0}^{(6)}(\lambda)}{V_{0}^{(2)}(\lambda)}=
-\frac{f_{c}^{2}}{2\eta_c^{2}r_{c}^{4}}
\left[-272f_{c}^{3}+408r_{c}^{2}f_{c}^{2}f_{c}^{(2)}
-88r_{c}^{3}f_{c}^{2}f_{c}^{(3)}\phantom{\left(f_{c}^{(2)}\right)^{2}}\right.
\nonumber
\\ &&  \qquad  \left.+r_{c}^{4}f_{c}\left(38f_{c}f_{c}^{(4)}
-204\left(f_{c}^{(2)}\right)^{2}\right)+r_{c}^{5}f_{c}\left(104f_{c}^{(2)}f_{c}^{(3)}
+18f_{c}f_{c}^{(5)}\right)
\phantom{\left(f_{c}^{(2)}\right)^{2}}\right. \nonumber
\\ && \qquad  \left.+r_{c}^{6}\left(34\left(f_{c}^{(2)}\right)^{3}+15f_{c}\left(f_{c}^{(3)}\right)^{2}
+26f_{c}f_{c}^{(2)}f_{c}^{(4)}+f_{c}^{2}f_{c}^{(6)}\right)\right]
+\underset{|\lambda| \to +\infty}{\cal O}\left(
\frac{1}{\lambda^2}\right).
\end{eqnarray}
\end{subequations}
\end{widetext}

We can now solve Eq.~(\ref{WKB_RP_1}) by
assuming $|\lambda| \gg 1$ as well as $\mathrm{Re} \, \lambda \gg
\mathrm{Im} \, \lambda$. We obtain for the solutions a family
$\lambda_n(\omega)$ with $n=1,2,3, \dots $ given by the
approximation
\begin{widetext}
\begin{equation}\label{PRapproxWKB}
\lambda_{n}(\omega) \approx \left[\frac{r_{c}^{2}}{f_{c}} \,
\omega^{2}
+a_{n}+2\eta_c^2\alpha(n)^{2}\epsilon_{n}(\omega)\right]^{1/2}+i\eta_c\alpha(n)\left[1+\epsilon_{n}(\omega)\right]
\end{equation}
where
\begin{eqnarray}\label{an}
&& a_{n}=-~\frac{1}{1152\eta_c^{4}}  \left\{
288f_{c}^{2}\left[(d^2-2d-1)f_{c}-(d-3)^{2}\right]\phantom{\left(f_{c}^{(2)}\right)^{3}}
\right. \nonumber \\ && \left. \quad +144r_{c}^{2}f_{c}f_{c}^{(2)}\left[2(d-3)^{2}-(2d^2-4d-3)f_{c}\right]
\phantom{\left(f_{c}^{(2)}\right)^{3}}\right.\nonumber \\
&&\left.\quad- 72r_{c}^{3}f_{c}^{2}f_{c}^{(3)} -18
r_{c}^{4}\left[4(d-3)^{2} \left(f_{c}^{(2)}\right)^{2} -4(d-3)(d+1)
f_{c}\left(f_{c}^{(2)}\right)^{2} + f_{c}^{2}f_{c}^{(4)}\right]\right.\nonumber \\
&&\left.\quad +
36r_{c}^{5}f_{c}f_{c}^{(2)}f_{c}^{(3)} + \,
r_{c}^{6}\left[36\left(f_{c}^{(2)}\right)^{3}-7f_{c}\left(f_{c}^{(3)}\right)^{2}
+9f_{c}f_{c}^{(2)}f_{c}^{(4)}\right] \right\} \nonumber \\
&&  \quad +~\alpha(n)^{2}\frac{r_c^3f_c}{96\eta_c^{4}} \left\{24f_cf_c^{(3)}+6r_cf_cf_c^{(4)}-12r_c^2f_c^{(2)}f_c^{(3)}+r_c^3\left(5\left(f_c^{(3)}\right)^2-3f_c^{(2)}f_c^{(4)}\right)\right\} \nonumber \\
&&
\end{eqnarray}
and
\begin{equation}\label{Epsilon}
\epsilon_{n}(\omega)=\frac{b_{n}}{(r_{c}^{2}/f_{c}) \, \omega^{2}+a_{n}+\eta_c^2\alpha(n)^{2}}
\end{equation}
with
\begin{eqnarray}\label{bn}
&& b_{n}=\frac{1}{442368\eta_c^{10}} \left\{
 -55296(d-3)^{2}f_{c}^{5}(1+f_{c})
+27648r_{c}^{2}f_{c}^{4}f_{c}^{(2)}\left[5(d-3)^{2}+2(d^2-12d+23)f_{c}\right]
\right. \nonumber \\ && \left. \quad - 9216 (3d^2-6d -2)
r_{c}^{3}f_{c}^{5}f_{c}^{(3)}
+6912r_{c}^{4}f_{c}^{3}\left[-20(d-3)^{2}\left(f_{c}^{(2)}\right)^{2}
+2(2d^2+36d-95)f_{c}\left(f_{c}^{(2)}\right)^{2}\right. \right. \nonumber \\ &&
\left. \left. \quad +9f_{c}^{2}f_{c}^{(4)} \right]
+3456r_{c}^{5}f_{c}^{4}\left[6(2d^2-4d+1)f_{c}^{(2)}f_{c}^{(3)}+5f_{c}f_{c}^{(5)}\right]
\right. \nonumber \\ && \left. \quad
-192r_{c}^{6}f_{c}^{2}\left[-360(d-3)^2\left(f_{c}^{(2)}\right)^{3}+144(2d^2+6d-25)
f_{c}\left(f_{c}^{(2)}\right)^{3}
+423f_{c}^{2}f_{c}^{(2)}f_{c}^{(4)}
-200f_{c}^{2}\left(f_{c}^{(3)}\right)^{2}\right. \right. \nonumber \\ &&\left. \left.\quad -5f_{c}^{3}f_{c}^{(6)}
\right]
-96r_{c}^{7}f_{c}^{3}\left[36(6d^2-12d+17)\left(f_{c}^{(2)}\right)^{2}f_{c}^{(3)}
-257f_{c}f_{c}^{(3)}f_{c}^{(4)}+270f_{c}f_{c}^{(2)}f_{c}^{(5)}\right]
\right. \nonumber \\ && \left. \quad
+12r_{c}^{8}f_{c}\left[-1440(d-3)^2\left(f_{c}^{(2)}\right)^{4}
+288(7d^2+6d-55)f_{c}\left(f_{c}^{(2)}\right)^{4}
+2376f_{c}^{2}\left(f_{c}^{(2)}\right)^{2}f_{c}^{(4)}\right. \right.
\nonumber \\ && \left. \left.   \qquad \qquad \quad +
67f_{c}^{3}\left(f_{c}^{(4)}\right)^{2}+152f_{c}^{3}
f_{c}^{(3)}f_{c}^{(5)}
 -2744f_{c}^{2}f_{c}^{(2)}\left(f_{c}^{(3)}\right)^{2}-120 f_{c}^{3}f_{c}^{(2)}f_{c}^{(6)} \right]
\right. \nonumber \\ &&   \left. \quad
+48r_{c}^{9}f_{c}^{2}\left[12(6d^2-12d+59)\left(f_{c}^{(2)}\right)^{3}f_{c}^{(3)}
+153f_{c}\left(f_{c}^{(3)}\right)^{3}-514f_{c}f_{c}^{(2)}f_{c}^{(3)}f_{c}^{(4)}
+270f_{c}\left(f_{c}^{(2)}\right)^{2}f_{c}^{(5)}\right]  \right.
\nonumber \\ && \left. \quad + 12 r_{c}^{10}\left[
144(d-3)^{2}\left(f_{c}^{(2)}\right)^{5}-288(d^2-7)
f_{c}\left(f_{c}^{(2)}\right)^{5}  -
67f_{c}^{3}f_{c}^{(2)}\left(f_{c}^{(4)}\right)^{2}-152f_{c}^{3}f_{c}^{(2)}f_{c}^{(3)}f_{c}^{(5)}
\right. \right. \nonumber \\ && \left. \left.   \qquad \qquad +
 344f_{c}^{2}\left(f_{c}^{(2)}\right)^{2}\left(f_{c}^{(3)}\right)^{2}
+60f_{c}^{3}\left(f_{c}^{(2)}\right)^{2}f_{c}^{(6)}
+108f_{c}^{2}\left(f_{c}^{(2)}\right)^{3}f_{c}^{(4)}
+153f_{c}^{3}\left(f_{c}^{(3)}\right)^{2}f_{c}^{(4)}\right]  \right.
\nonumber \\ && \left. \quad -24
r_{c}^{11}f_{c}f_{c}^{(2)}\left[252\left(f_{c}^{(2)}\right)^{3}f_{c}^{(3)}
+153f_{c}\left(f_{c}^{(3)}\right)^{3}-257f_{c}f_{c}^{(2)}f_{c}^{(3)}f_{c}^{(4)}
+90f_{c}\left(f_{c}^{(2)}\right)^{2}f_{c}^{(5)}\right]  \right.
\nonumber \\ && \left. \quad
+r_{c}^{12}\left[-864\left(f_{c}^{(2)}\right)^{6}+385f_{c}^{2}\left(f_{c}^{(3)}\right)^{4}
-1512f_{c}\left(f_{c}^{(2)}\right)^{4}f_{c}^{(4)}
-918f_{c}^{2}f_{c}^{(2)}\left(f_{c}^{(3)}\right)^{2}f_{c}^{(4)}
\right. \right. \nonumber \\ && \left. \left.   \qquad \quad +
201f_{c}^{2}\left(f_{c}^{(2)}\right)^{2}\left(f_{c}^{(4)}\right)^{2}
+456f_{c}^{2}\left(f_{c}^{(2)}\right)^{2}f_{c}^{(3)}f_{c}^{(5)} +
1368f_{c}\left(f_{c}^{(2)}\right)^{3}\left(f_{c}^{(3)}\right)^{2}
-120f_{c}^2 \left(f_{c}^{(2)}\right)^{3}f_{c}^{(6)} \right] \right\}
\nonumber \\ &&
 \quad +\frac{\alpha(n)^{2}r_{c}^{3}f_{c}}{110592
\eta_c^{10}}\left\{
 9216f_{c}^{4}f_{c}^{(3)}+13824r_{c}f_{c}^{4}f_{c}^{(4)}
+3456r_{c}^{2}f_{c}^{3}\left[-2f_{c}^{(2)}f_{c}^{(3)}+f_{c}f_{c}^{(5)}\right]
\phantom{\left(f_{c}^{(3)}\right)^{2}} \right. \nonumber \\ &&
\left. \quad
+192r_{c}^{3}f_{c}^{3}\left[72\left(f_{c}^{(3)}\right)^{2}
-99f_{c}^{(2)}f_{c}^{(4)}+f_{c}f_{c}^{(6)}\right]
-288r_{c}^{4}f_{c}^{2}\left[12\left(f_{c}^{(2)}\right)^{2}f_{c}^{(3)}
-29f_{c}f_{c}^{(3)}f_{c}^{(4)}+18f_{c}f_{c}^{(2)}f_{c}^{(5)}\right]
\right. \nonumber \\ && \left. \quad
+12r_{c}^{5}f_{c}^{2}\left[648\left(f_{c}^{(2)}\right)^{2}f_{c}^{(4)}
+ 17f_{c}\left(f_{c}^{(4)}\right)^{2}+56f_{c}f_{c}^{(3)}f_{c}^{(5)}
-
1032f_{c}^{(2)}\left(f_{c}^{(3)}\right)^{2}-24f_{c}f_{c}^{(2)}f_{c}^{(6)}
\right] \right. \nonumber \\ && \left. \quad
+144r_{c}^{6}f_{c}\left[28\left(f_{c}^{(2)}\right)^{3}f_{c}^{(3)}
+25f_{c}\left(f_{c}^{(3)}\right)^{3}-58f_{c}f_{c}^{(2)}f_{c}^{(3)}f_{c}^{(4)}
+18f_{c}\left(f_{c}^{(2)}\right)^{2}f_{c}^{(5)}\right]  \right.
\nonumber \\ && \left. \quad
+12r_{c}^{7}f_{c}\left[-36\left(f_{c}^{(2)}\right)^{3}f_{c}^{(4)}
+75f_{c}\left(f_{c}^{(3)}\right)^{2}f_{c}^{(4)}-
17f_{c}f_{c}^{(2)}\left(f_{c}^{(4)}\right)^{2}
-56f_{c}f_{c}^{(2)}f_{c}^{(3)}f_{c}^{(5)} \right. \right.   \nonumber \\
&& \left. \left.  \qquad \qquad \quad +
168\left(f_{c}^{(2)}\right)^{2}\left(f_{c}^{(3)}\right)^{2} +12
f_{c}\left(f_{c}^{(2)}\right)^{2}f_{c}^{(6)} \right]\right.
\nonumber
\\ && \left. \quad
-72r_{c}^{8}f_{c}^{(2)}\left[12\left(f_{c}^{(2)}\right)^{3}f_{c}^{(3)}
+25f_{c}\left(f_{c}^{(3)}\right)^{3}-29f_{c}f_{c}^{(2)}f_{c}^{(3)}f_{c}^{(4)}
+6f_{c}\left(f_{c}^{(2)}\right)^{2}f_{c}^{(5)}\right]\right.
\nonumber \\ && \left. \quad
+r_{c}^{9}\left[235f_{c}\left(f_{c}^{(3)}\right)^{4}-216\left(f_{c}^{(2)}\right)^{4}f_{c}^{(4)}
-450f_{c}f_{c}^{(2)}\left(f_{c}^{(3)}\right)^{2}f_{c}^{(4)}   +
51f_{c}\left(f_{c}^{(2)}\right)^{2}\left(f_{c}^{(4)}\right)^{2}
\right. \right. \nonumber \\ && \left. \left.   \qquad
+168f_{c}\left(f_{c}^{(2)}\right)^{2}f_{c}^{(3)}f_{c}^{(5)} +
 360\left(f_{c}^{(2)}\right)^{3}\left(f_{c}^{(3)}\right)^{2}
-24f_{c}\left(f_{c}^{(2)}\right)^{3}f_{c}^{(6)} \right] \right\}.
\end{eqnarray}
\end{widetext}
Equation (\ref{PRapproxWKB}) is the main result of our paper. As we
shall see below (see also Appendix A), it also provides expressions
for the dispersion relation and the damping of the ``surface waves"
lying on (close to) the photon sphere of the considered BH.

It is moreover possible to simplify (\ref{PRapproxWKB}) and to
obtain a high frequency approximation for the Regge poles. We have
\begin{widetext}
\begin{equation}\label{RPHF}
\lambda_{n}(\omega)=\left[\frac{r_{c}}{\sqrt{f_{c}}}~\omega
+\frac{a_{n}}{(2r_{c}/\sqrt{f_{c}})}~\frac{1}{\omega}\right]
+i\eta_c\alpha(n)\left[1+\frac{b_n}{(r_{c}^{2}/f_{c})}~\frac{1}{\omega^{2}}\right]
+\underset{\omega \to +\infty}{\cal O}\left(
\frac{1}{\omega^3}\right).
\end{equation}
\end{widetext}
It is important to understand the difference between the
approximation (\ref{PRapproxWKB}) and its much more elegant version
(\ref{RPHF}). The approximation (\ref{RPHF}) is meaningful as a
$1/\omega$ expansion with $\omega \to +\infty$. By contrast, the
approximation (\ref{PRapproxWKB}) remains valid in a large range of
frequencies thanks to WKB theory. Indeed, in order to establish
Eq.~(\ref{WKB_RP_1}) we have considered as a perturbation parameter
of the WKB method the distance between the turning points of
Eq.~(\ref{RW}) [i.e., the roots of $Q_\ell(r^\ast)= \omega^2 -
V_{\ell}(r^\ast)$] and the location of the peak of $Q_\ell(r^\ast)$
(see also Refs.~\cite{BenderOrszag1978,SchutzWill,Iyer1,Iyer2})
instead of $1/\omega$. Of course, in order to solve
Eq.~(\ref{WKB_RP_1}) and to obtain the expression
(\ref{PRapproxWKB}), we have furthermore assumed that $|\lambda| \gg
1$ as well as $\mathrm{Re} \, \lambda \gg \mathrm{Im} \, \lambda$
and, as a consequence, we cannot expect from (\ref{PRapproxWKB}) a
very high accuracy for very low frequencies (see also
Refs.~\cite{DolanOttewill_2009} and \cite{DecaniniFolacci2010a} for
related numerical studies in the case of the Schwarzschild BH).

It is furthermore interesting to provide an interpretation of the
previous results in terms of ``surface waves" as it is customary in
the CAM approach. As we have noted in Appendix A, the resonant part
of the form factor is a superposition of terms like
$\exp[i(\lambda_n(\omega)\theta_1-\omega t)]$ (in this paragraph we
take into account the harmonic time dependence $\exp[-i\omega t]$).
By inserting the leading-order terms of Eq.~(\ref{RPHF}) into these
wavelike contributions, we can easily note that the contribution of
the $n$th ``surface wave" reduces to
\begin{equation}\label{surface_wave}
\exp[-\eta_c \alpha(n) \theta_1] \, \exp\left[i \left(r_c/\sqrt{f_c}
\,\omega \theta_1-\omega t \right)\right]
\end{equation}
The second term describes the propagation of this ``surface wave"
near the photon sphere at $r=r_c$. Indeed, it circles the BH in time
$T'=2\pi (r_c/\sqrt{f_c})$ which is exactly the time (\ref{Torbit})
needed for a massless particle to orbit the BH on an unstable
circular null geodesic. The first term corresponds to an exponential
decay of this ``surface wave" due to continual reradiation of
energy. It is very interesting to rewrite Eq.~(\ref{surface_wave})
in the form
\begin{equation}\label{Interp_SW}
\exp[-k''_n(\omega){\cal L}]\exp[i(k'_n(\omega){\cal L}-\omega t)]
\end{equation}
with ${\cal L}=r_c \, \theta_1$ which denotes the arc length taken
on the photon sphere. Now, $k'_n(\omega)=\omega/\sqrt{f_c}$
represents the wavenumber of the $n$th ``surface wave" or, in other
terms, its dispersion relation, while $k''_n(\omega)=\eta_c
\alpha(n)/r_c$ is its damping constant.

To the leading order, the dispersion relation is linear and
independent of the index of the ``surface wave" while the damping
constant depends only on the index $n$. Of course, if we go beyond
the leading-order terms, the dispersion relation reads [see
Eq.~(\ref{PRapproxWKB})]
\begin{subequations}\label{PR_WKB_disp}
\begin{equation}\label{PRapproxWKB_disp}
k'_{n}(\omega) \approx \left[\frac{1}{f_{c}} \,\omega^{2} +
\frac{a_{n}}{r_c^2}+2\left(\frac{\eta_c^2}{r_c^2
}\right)\alpha(n)^{2}\epsilon_{n}(\omega)\right]^{1/2}
\end{equation}
which implies [see Eq.~(\ref{RPHF})]
\begin{equation} \label{PRasympWKB_disp}
k'_{n}(\omega)=\left[\frac{1}{\sqrt{f_{c}}}~\omega
+\frac{a_{n}}{(2r_c^2/\sqrt{f_{c}})}~\frac{1}{\omega}\right]
+\underset{\omega \to +\infty}{\cal O}\left(
\frac{1}{\omega^3}\right)
\end{equation}
\end{subequations}
and is clearly nonlinear as well as dependent on the index $n$. This
result could have important consequences in strong gravitational
lensing (see also Ref.~\cite{DecaniniFolacci2010a} and the
discussion in Appendix B.1 of the present paper). With such a
potential application in mind, it is worth noting that, in
Eqs.~(\ref{Interp_SW}) and (\ref{PR_WKB_disp}), $\omega$ denotes the
frequency of the scalar photon observed by a static observer at
infinity. If we consider the gravitational redshift of this photon
and introduce its frequency ${\tilde \omega}=\omega/\sqrt{f_{c}}$
measured by a static observer lying on the photon sphere, the
dispersion relation (\ref{PR_WKB_disp}) reads
\begin{subequations}\label{PR_WKB_disp_GS}
\begin{equation}\label{PRapproxWKB_disp_GS}
k'_{n}({\tilde \omega}) \approx \left[{\tilde \omega}^{2}
+\frac{a_{n}}{r_c^2}+2\left(\frac{\eta_c^2}{r_c^2
}\right)\alpha(n)^{2}\epsilon_{n}({\tilde \omega})\right]^{1/2}
\end{equation}
and we have
\begin{equation}\label{RPHF_disp_GS} k'_{n}({\tilde
\omega})=\left[{\tilde \omega}
+\frac{a_{n}}{2r_c^2}~\frac{1}{{\tilde \omega}}\right]
+\underset{\omega \to +\infty}{\cal O}\left( \frac{1}{{\tilde
\omega}^3}\right).
\end{equation}
\end{subequations}
When $a_n>0$, Eq.~(\ref{PR_WKB_disp_GS}) provides a superluminal
dispersion relation. Indeed, it leads to a group velocity
$v_g({\tilde \omega})=d{\tilde \omega} /dk'_{n}({\tilde \omega})
>1$. The condition $a_n>0$ is satisfied $\forall n \in {\bf N}$ for
spins 1 and 2 propagating on the Schwarzschild BH [see Eq.~(13) of
Ref.~\cite{DecaniniFolacci2010a}]. It is also satisfied for the
massless scalar field when $n\not= 0$. As a consequence, it cannot
be considered as an exotic physical condition. It even seems to be
true in most cases encountered (see the examples considered in
Sec.~IV). At first sight, this result may seem a little bit
puzzling. But it is important to note that superluminal dispersion
relations for the ``surface waves" lying on the photon sphere do not
necessarily lead to a violation of the relativistic principle of
causality. Indeed, for the transfer of information between a source
and a receptor located outside a BH, various ``channels" are
involved. Of course, there are channels associated with diffraction
by the BH photon sphere [for a source and a receptor at infinity,
they correspond to the sum over the Regge poles in
Eq.~(\ref{ampliPR2})] but there are also channels associated with
geometrical rays [for a source and a receptor at infinity, they come
from the background integral (\ref{SWTransfo}) over the contour
$\Gamma$ of Fig.~\ref{fig:watson} after asymptotic evaluation]. In
order to study causality, it would be necessary to carefully take
into account interferences between all the monochromatic spectral
components of the signal carrying the information for all the
channels involved. We believe that such a study would not show a
violation of causality. The situation encountered here is similar to
that discussed
 by many authors working in electromagnetism of dispersive media
 (see, e.g., Ref.~\cite{Brillouin1960} and references therein). In such a context, it
  has been observed that a group velocity greater than the light velocity does not violate
   causality because it is not the velocity of information transmission or the energy velocity.

Finally, the WKB result (\ref{PRapproxWKB}) and the associated Regge
trajectories permit us to derive, from the semiclassical formulas
(\ref{sc1}) and (\ref{sc4}), useful analytical expressions for the
QNM complex frequencies. Indeed, by inserting (\ref{PRapproxWKB})
into Eqs.~(\ref{sc1}) and (\ref{sc4}), we obtain the large $\ell$
behaviors
\begin{widetext}
\begin{subequations}\label{asympQNM}
\begin{eqnarray}
& & \omega ^{(o)}_{\ell n} = \frac{\sqrt{f_{c}}}{r_{c}} \left[ \left(\ell +\frac{d-3}{2}\right)
-  \frac{a_n}{2 \ell} +  (d-3)\frac{a_n}{4 \ell^2} + \underset{\ell \to
+\infty}{\cal O}\left( \frac{1}{\ell^3} \right)
 \right], \\
 & & \frac{\Gamma _{\ell n}}{2} =
\eta_c\frac{\sqrt{f_{c}}}{r_{c}} \alpha(n)\left[1+ \frac{c_n}{2 \ell^2} +
\underset{\ell \to +\infty}{\cal O}\left( \frac{1}{\ell^3}
\right)\right],
\end{eqnarray}
\end{subequations}
with $\ell \in {\bf N}$ and $n=1, 2, \dots$ Here
\begin{equation}\label{c_n}
c_n=a_n+2 b_n.
\end{equation}
\end{widetext}
It should be noted that the leading-order terms of
Eq.~(\ref{asympQNM}) have already been obtained in
Refs.~\cite{ZerbiniVanzo2004,CardosoMirandaBertietal2009,DolanOttewill_2009}.
The higher-order terms in $1/\ell$ and $1/\ell^2$ are new (see,
however, Sec.~5.2 of Ref.~\cite{DolanOttewill_2009}). They are
directly linked to the nonlinear behavior of both the dispersion
relation and the damping of the ``surface waves" lying on (close to)
the photon sphere.

\section{Applications}

In this section we shall apply the previous formalism to various
spacetimes of physical interest.

\subsection{Schwarzschild-Tangherlini black holes}
For Schwarzschild-Tangherlini BHs \cite{Tangherlini63}, the function $f(r)$ reads
\begin{equation}
f(r)=1-\left(\frac{r_0}{r}\right)^{d-3}
\end{equation}
with
\begin{equation}\label{r0dimd}
r_0^{d-3}=\frac{16\pi M}{(d-2){\cal A}_{d-2}} \quad \text{and} \quad
{\cal A}_{d-2}=\frac{2\pi^{(d-1)/2}}{\Gamma[(d-1)/2]}.
\end{equation}
Here $M$ is the mass of the BH, ${\cal A}_{d-2}$ the area of the
unit sphere $S^{d-2}$ and the event horizon is located at $r_h=r_0$.
The unstable circular null geodesics are located at
\begin{subequations}
\begin{equation}\label{rcSchTang}
r_c=r_0\left(\frac{d-1}{2}\right)^{1/(d-3)}
\end{equation}
and the associated $\eta_c$ parameter is given by
\begin{equation}\label{etaSchTang}
\eta_c=\sqrt{d-3}.
\end{equation}
\end{subequations}

\subsubsection{d=4. The Schwarzschild black hole}

For $d=4$, the Schwarzschild-Tangherlini solution is nothing but the
ordinary Schwarzschild BH and we have
\begin{subequations}
\begin{eqnarray}
&&r_{c}=\frac{3}{2}r_{0}=3M,\\
&&\eta_c=1,
\end{eqnarray}
\end{subequations}
as well as
\begin{subequations}
\begin{eqnarray}
&&a_{n}=-\frac{29}{216}+\frac{5}{18}\alpha(n)^{2},\\
&&b_{n}=\frac{1357}{15552}-\frac{305}{3888}\alpha(n)^{2},\\
&&c_{n}=\frac{313}{7776}+\frac{235}{1944}\alpha(n)^{2},\\
&&\epsilon_{n}(\omega)=\frac{b_{n}}{27M^{2}\omega^{2}+a_{n}+\alpha(n)^{2}}.
\end{eqnarray}
\end{subequations}
The WKB approximation (\ref{PRapproxWKB}) for the Regge poles leads to
\begin{eqnarray}\label{Sch41}
&&\lambda_{n}(\omega) \approx
\left[27M^{2}\omega^{2}+a_{n}+2\alpha(n)^{2}\epsilon_{n}(\omega)\right]^{1/2}
\nonumber \\ && \qquad+i\alpha(n)\left[1+\epsilon_{n}(\omega)\right]
\end{eqnarray}
and their high frequency behavior (\ref{RPHF}) provides
\begin{eqnarray}\label{Sch42}
&&\lambda_{n}(\omega)=\left[3\sqrt{3}M\omega+\frac{\sqrt{3}a_n}{18M\omega}\right]\nonumber \\
&& \qquad+i\alpha(n)\left[1+\frac{b_{n}}{27M^{2}\omega^{2}}\right]
+\underset{\omega \to +\infty}{\cal O}\left(
\frac{1}{\omega^3}\right).
\end{eqnarray}
Formulas (\ref{Sch41}) and (\ref{Sch42}) are in agreement with the results
obtained in Ref.~\cite{DecaniniFolacci2010a}.

The resonance excitation frequencies and the damping of the QNMs given by
the general formulas (\ref{asympQNM}) reduce to
\begin{subequations}\label{QNMSch}
\begin{eqnarray}
&&\omega_{\ell
n}^{(0)}=\frac{1}{3\sqrt{3}M}\left[\left(\ell+\frac{1}{2}
\right)-\frac{a_{n}}{2\ell} +\frac{a_{n}}{4\ell^{2}}+\underset{\ell
\to +\infty}{\cal O}\left(\frac{1}{\ell^3}\right)\right],
\nonumber \\
&&\\
&&\frac{\Gamma_{\ell n}}{2}=\frac{\alpha(n)}{3\sqrt{3}M}\left[1+\frac{c_{n}}{2\ell^{2}}
+\underset{\ell \to +\infty}{\cal O}\left(
\frac{1}{\ell^3}\right)\right].
\end{eqnarray}
\end{subequations}
Formulas (\ref{QNMSch}) are in agreement with the results obtained in Ref.~\cite{Iyer2}
(see also Ref.~\cite{DecaniniFolacci2010a}).

\subsubsection{The five-dimensional Schwarzschild-Tangherlini black hole}
For $d=5$, we have
\begin{subequations}
\begin{eqnarray}
&&r_{c}=\sqrt{2}r_{0},\\
&&\eta_c=\sqrt{2},
\end{eqnarray}
\end{subequations}
as well as
\begin{subequations}
\begin{eqnarray}
&&a_{n}=-\frac{5}{16}+\frac{3}{4}\alpha(n)^{2},\\
&&b_{n}=\frac{115}{512}-\frac{31}{128}\alpha(n)^{2},\\
&&c_{n}=\frac{5}{48}+\frac{17}{64}\alpha(n)^{2},\\
&&\epsilon_{n}(\omega)=\frac{b_{n}}{2r_{c}^{2} \, \omega^{2}+a_{n}+2\alpha(n)^{2}}.
\end{eqnarray}
\end{subequations}
The WKB approximation (\ref{PRapproxWKB}) for the Regge poles leads to
\begin{eqnarray}
&&\lambda_{n}(\omega) \approx
\left[2r_{c}^{2}\omega^{2}+a_{n}+4\alpha(n)^{2}\epsilon_{n}(\omega)\right]^{1/2}\nonumber
\\ && \qquad+i\alpha(n)[1+\epsilon_{n}(\omega)]
\end{eqnarray}
and their high frequency behavior (\ref{RPHF}) provides
\begin{eqnarray}
&&\lambda_{n}(\omega)=\left[\sqrt{2}\,r_{c}\omega+\frac{a_{n}}{2\sqrt{2}\,r_{c}\omega}\right]\nonumber \\ &&\qquad+i\sqrt{2}\,\alpha(n)\left[1+\frac{b_n}{2r_{c}^{2}\omega^{2}}\right]+\underset{\omega \to +\infty}{\cal O}\left(
\frac{1}{\omega^3}\right).
\end{eqnarray}

The resonance excitation frequencies and the damping of the QNMs given
by the general formulas (\ref{asympQNM}) reduce to
\begin{subequations}
\begin{eqnarray}
&&\omega_{\ell n}^{(0)}=\frac{1}{\sqrt{2}\,r_c}\left[\left(\ell+1\right)-\frac{a_{n}}{2\ell}+\frac{a_{n}}{2\ell^{2}}+\underset{\ell \to +\infty}{\cal O}\left(
\frac{1}{\ell^3}\right)\right],\nonumber \\
&&\\
&&\frac{\Gamma_{\ell n}}{2}=\frac{\alpha(n)}{r_c}\left[1+\frac{c_{n}}{2\ell^{2}}+\underset{\ell \to +\infty}{\cal O}\left(
\frac{1}{\ell^3}\right)\right].
\end{eqnarray}
\end{subequations}

\subsubsection{The six-dimensional Schwarzschild-Tangherlini black hole}
For $d=6$, we have
\begin{subequations}
\begin{eqnarray}
&&r_{c}=\left(\frac{5}{2}\right)^{1/3}r_{0},\\
&&\eta_c=\sqrt{3},
\end{eqnarray}
\end{subequations}
as well as
\begin{subequations}
\begin{eqnarray}
&&a_{n}=-\frac{31}{60}+\frac{7}{5}\alpha(n)^{2},\\
&&b_{n}=\frac{1411}{3600}-\frac{91}{180}\alpha(n)^{2},\\
&&c_{n}=\frac{481}{1800}+\frac{7}{18}\alpha(n)^{2},\\
&&\epsilon_{n}(\omega)=\frac{b_{n}}{(5/3)r_{c}^{2} \, \omega^{2}+a_{n}
+3\alpha(n)^{2}}.
\end{eqnarray}
\end{subequations}
The WKB approximation (\ref{PRapproxWKB}) for the Regge poles leads to
\begin{eqnarray}
&&\lambda_{n}(\omega) \approx \left[\frac{5}{3}r_{c}^{2}\omega^{2}
+a_{n}+6\alpha(n)^{2}\epsilon_{n}(\omega)\right]^{1/2}\nonumber \\
&& \qquad+i\sqrt{3}\alpha(n)\left[1+\epsilon_{n}(\omega)\right]
\end{eqnarray}
and their high frequency behavior (\ref{RPHF}) provides
\begin{eqnarray}
&&\lambda_{n}(\omega)=\left[\left(\frac{5}{3}\right)^{1/2}r_{c}\omega+\frac{a_{n}}{2(5/3)^{1/2}r_{c}\omega}\right]\nonumber \\ && +i\sqrt{3}\,\alpha(n)\left[1+\frac{b_{n}}{(5/3)r_{c}^{2} \, \omega^{2}}\right]+\underset{\omega \to +\infty}{\cal O}\left(
\frac{1}{\omega^3}\right).\nonumber \\
&&
\end{eqnarray}

The resonance excitation frequencies and the damping of the QNMs given by the general formulas (\ref{asympQNM}) reduce to
\begin{subequations}
\begin{eqnarray}
&&\omega_{\ell
n}^{(0)}=\left(\frac{3}{5}\right)^{1/2}\left(\frac{1}{r_{c}}\right)
\left[\left(\ell+ \frac{3}{2} \right)\phantom{\underset{\ell \to
+\infty}{\cal O}\left( \frac{1}{\ell^3}\right)}\right.\nonumber\\ &&
\qquad \quad \left. -\frac{a_{n}}{2\ell}+3\frac{a_{n}}{4\ell^{2}}
+\underset{\ell \to +\infty}{\cal O}\left(
\frac{1}{\ell^3}\right)\right],\\
&&\frac{\Gamma_{\ell n}}{2}=
\sqrt{3}\left(\frac{3}{5}\right)^{1/2}\left(\frac{1}{r_{c}}\right)\alpha(n)\nonumber\\
&& \qquad \quad \times \left[1+\frac{c_{n}}{2\ell^{2}}+\underset{\ell \to +\infty}{\cal O}\left(
\frac{1}{\ell^3}\right)\right].
\end{eqnarray}
\end{subequations}

\subsubsection{Leading-order terms for the Schwarzschild-Tangherlini black hole of arbitrary dimension}

It should be noted that, in the $d$-dimensional case, we can easily
derive the leading-order terms of the Regge poles and quasinormal
complex frequencies. We have
\begin{eqnarray}
&&\lambda_{n}(\omega)=\sqrt{\frac{d-1}{d-3}}\,r_c\omega+i\sqrt{d-3}\,\alpha(n)+\underset{\omega \to +\infty}{\cal O}\left(\frac{1}{\omega}\right)\nonumber\\
&&
\end{eqnarray}
and
\begin{subequations}\label{QNMSchTang}
\begin{eqnarray}
&&\omega_{\ell n}^{(0)} = \sqrt{\frac{d-3}{d-1}}\left(\frac{1}{r_{c}}\right)\left(\ell+\frac{d-3}{2}\right)+\underset{\ell \to +\infty}{\cal O}\left(
\frac{1}{\ell}\right),\nonumber \\
&&\\
&&\frac{\Gamma_{\ell n}}{2} = \frac{d-3}{\sqrt{d-1}}\left(\frac{1}{r_{c}}\right)\alpha(n)+\underset{\ell \to +\infty}{\cal O}\left(\frac{1}{\ell}\right).
\end{eqnarray}
\end{subequations}
Formulas (\ref{QNMSchTang}) are in agreement with the results
obtained in Refs.~\cite{Konoplya2003} and \cite{ZerbiniVanzo2004}.

\subsection{The Reissner-Nordstr\"om black hole}
For the $d$-dimensional Reissner-Nordstr\"om BH \cite{Tangherlini63}, the function $f(r)$ reads
\begin{equation}
f(r)=1-\left(\frac{r_0}{r}\right)^{d-3}+\frac{\theta^{2}}{r^{2(d-3)}}
\end{equation}
with (see Ref.~\cite{MyersPerry1986} or Appendix A of Ref.~\cite{NatarioSchiappa2004})
\begin{equation}
r_{0}^{d-3}=\frac{16\pi M}{(d-2){\cal A}_{d-2}}\doteq 2\mu \quad \text{and}
\quad \theta^2=\frac{8\pi q^2}{(d-2)(d-3)}.
\end{equation}
Here $M$ is the mass of the BH and $q$ is its charge. We furthermore
assume that $\mu^2>\theta^2$. It should be noted that here we
consider electromagnetism in the Heaviside system of units. For this
background there are two horizons, the so-called inner and outer
horizons. They are respectively located at
\begin{subequations}
\begin{eqnarray}
&&r_{-}=\left(\mu-\sqrt{\mu^{2}-\theta^{2}}\right)^{1/(d-3)},\\
&&r_{+}=\left(\mu+\sqrt{\mu^{2}-\theta^{2}}\right)^{1/(d-3)}.
\end{eqnarray}
\end{subequations}
We are only interested in the outer horizon with radius at
$r_h=r_{+}$ because we have $f(r)>0$ for $r \in ]r_h,+\infty[$ [see
assumption (i) of Sec.~II].

The unstable circular null geodesics are located at $r_c>r_h$ given
by
\begin{subequations}
\begin{equation}
r_c=\left(\frac{(d-1)\mu+\sqrt{[(d-1)\mu]^{2}-4(d-2)\theta^{2}}}{2}\right)^{1/(d-3)},
\end{equation}
and we have for the associated parameter $\eta_c$
\begin{equation}
\eta_c=\sqrt{(d-3)-\frac{(d-2)(d-3)\theta^2}{r_c^{2(d-3)}}}.
\end{equation}
\end{subequations}
For $\theta=0$ we recover the Schwarzschild-Tangherlini BH results.
It should be noted that it is also possible to express $\eta_c$ in two other
equivalent forms since, as a consequence of Eq.~(\ref{Assup_fr_2a}),
the parameters $r_c$, $\mu$ and $\theta$ are linked by
\begin{equation}
\left(r_c^{d-3}\right)^{2}-(d-1)\mu r_c^{d-3} +(d-2)\theta^2=0.
\end{equation}

\subsubsection{The four-dimensional Reissner-Nordstr\"om black hole}

For $d=4$, we have
\begin{subequations}
\begin{eqnarray}
&&r_c=\frac{1}{2}(3M+\sqrt{9M^{2}-8Q^{2}}),\\
&&\eta_c=\sqrt{1-\frac{2Q^{2}}{r_c^{2}}},
\end{eqnarray}
\end{subequations}
as well as
\begin{widetext}
\begin{subequations}
\begin{eqnarray}
&&a_{n}=\frac{\left(1-Q^2/r_c^2\right)^2}{216(1-2Q^2/r_c^2)^2}
\left[-29+\frac{86Q^2}{r_c^2}\right]
+\frac{1-Q^2/r_c^2}{36(1-2Q^2/r_c^2)^2}\left[10-\frac{62Q^2}{r_c^2}
+\frac{124Q^4}{r_c^4}\right]\alpha(n)^{2},\\
&&b_{n}=\frac{\left(1-Q^2/r_c^2\right)^2}{15552(1-2Q^2/r_c^2)^5}\left[1357
-\frac{12478 Q^2}{r_c^2} + \frac{42333 Q^4}{r_c^4} -\frac{64324 Q^6}{r_c^6}
+ \frac{43588 Q^8}{r_c^8}\right]\nonumber\\
&& \quad +\frac{1-Q^2/r_c^2}{3888(1-2Q^2/r_c^2)^5}\left[-305 + \frac{3943 Q^2}{r_c^2} -
 \frac{21335 Q^4}{r_c^4} + \frac{56357 Q^6}{r_c^6} - \frac{69544 Q^8}{r_c^8}
 + \frac{34772 Q^{10}}{r_c^{10}}\right]\alpha(n)^{2},  \\
&&c_{n}=\frac{\left(1-Q^2/r_c^2\right)^2}{7776(1-2Q^2/r_c^2)^5}\left[313
- \frac{3118 Q^2}{r_c^2}
+ \frac{11229 Q^4}{r_c^4} - \frac{18820 Q^6}{r_c^6} + \frac{18820 Q^8}{r_c^8}\right]\nonumber\\
&& \quad +\frac{1-Q^2/r_c^2}{1944(1-2Q^2/r_c^2)^5}\left[235 -
\frac{2645 Q^2}{r_c^2} + \frac{11929 Q^4}{r_c^4} - \frac{28315
Q^6}{r_c^6} + \frac{37592 Q^8}{r_c^8} - \frac{18796
Q^{10}}{r_c^{10}}\right]\alpha(n)^{2},
\end{eqnarray}
and
\begin{eqnarray}
&&\epsilon_{n}(\omega)=\frac{b_{n}}{3r_c^2/(1-Q^2/r_c^2)\,\omega^{2}
+a_{n}+(1-2Q^{2}/r_c^{2})\alpha(n)^{2}}.
\end{eqnarray}
\end{subequations}
Here we have noted $Q^2=4\pi q^2$. This permits us to compare our
results with those for which electromagnetism is expressed in the
Gaussian system of units. The WKB approximation (\ref{PRapproxWKB})
for the Regge poles leads to
\begin{eqnarray}
&&\lambda_{n}(\omega) \approx \left[\frac{3r_c^2}{1-Q^2/r_c^2}
\omega^{2}+a_{n}
 +2\left(1-\frac{2Q^{2}}{r_c^{2}}\right)\alpha(n)^{2}\epsilon_{n}(\omega)\right]^{1/2}
+i\sqrt{1-\frac{2Q^{2}}{r_c^{2}}}\alpha(n)[1+\epsilon_{n}(\omega)]
\end{eqnarray}
\end{widetext}
and their high frequency behavior (\ref{RPHF}) provides
\begin{eqnarray}
&&\lambda_{n}(\omega)=\left[\frac{\sqrt{3}\,r_c}{\sqrt{1-Q^2/r_c^2}} \omega+\frac{a_{n}}{2\sqrt{3}\,r_c/\sqrt{1-Q^2/r_c^2} \,\omega}\right]\nonumber \\ && \quad +i\sqrt{1-\frac{2Q^{2}}{r_c^{2}}}\alpha(n)\left[1+\frac{b_n}{3r_c^2/(1-Q^2/r_c^2)\,\omega^{2}}\right]\nonumber \\ && \quad +\underset{\omega \to +\infty}{\cal O}\left(
\frac{1}{\omega^3}\right).
\end{eqnarray}

The resonance excitation frequencies and the damping of the QNMs given
by the general formulas (\ref{asympQNM}) reduce to
\begin{subequations}\label{QNMRN}
\begin{eqnarray}
&&\omega_{\ell
n}^{(0)}=\sqrt{1-\frac{Q^2}{r_c^2}}\left(\frac{1}{\sqrt{3}\,r_c}\right)
\left[\left(\ell+ \frac{1}{2} \right) \phantom{\left(
\frac{1}{\ell^3}\right)}\right. \nonumber \\ && \left.\qquad
-\frac{a_{n}}{2\ell}+\frac{a_{n}}{4\ell^{2}}+\underset{\ell \to
+\infty}{\cal O}\left(
\frac{1}{\ell^3}\right)\right],\\
&&\frac{\Gamma_{\ell n}}{2}= \sqrt{1-\frac{2Q^{2}}{r_c^{2}}}\sqrt{1-\frac{Q^2}{r_c^2}}
\left(\frac{1}{\sqrt{3}\,r_c}\right)\,\alpha(n) \nonumber \\ && \qquad \times
\left[1+\frac{c_{n}}{2\ell^{2}}+\underset{\ell \to +\infty}{\cal O}\left(
\frac{1}{\ell^3}\right)\right].
\end{eqnarray}
\end{subequations}
The leading-order terms of (\ref{QNMRN}) are in agreement with
formulas (37) and (38) of Ref.~\cite{FerrariMashhoon84}.
Furthermore, it should be noted that, for $Q=0$, all the results
obtained for the Regge poles and the complex quasinormal frequencies
of the four-dimensional Reissner-Nordstr\"om BH reduce to the
Schwarzschild BH results of Sec.~IV.A.1.

\subsubsection{Leading-order terms for the Reissner-Nordstr\"om black hole of arbitrary dimension}
In the $d$-dimensional case, we can easily derive the leading-order
terms of the Regge poles and quasinormal complex frequencies. We
have
\begin{eqnarray}\label{PR_RNd}
&&\lambda_{n}(\omega)=\sqrt{\frac{d-1}{d-3}}\frac{r_{c}\omega}{\sqrt{1-\frac{\theta^2}{r_c^{2(d-3)}}}
} \nonumber \\ && \qquad  +i\sqrt{d-3}\sqrt{1-\frac{(d-2)\theta^2}{r_c^{2(d-3)}}}\alpha(n)
+\underset{\omega \to +\infty}{\cal O}\left(\frac{1}{\omega}\right)\nonumber\\
&&
\end{eqnarray}
and
\begin{subequations}\label{QNMRNd}
\begin{eqnarray}
&&\omega_{\ell n}^{(0)} = \sqrt{\frac{d-3}{d-1}}\sqrt{1-\frac{\theta^2}{r_c^{2(d-3)}}}
\left(\frac{1}{r_{c}}\right)\left(\ell+\frac{d-3}{2}\right)\nonumber \\
&& \qquad +\underset{\ell \to +\infty}{\cal O}\left(
\frac{1}{\ell}\right),\\
&&\frac{\Gamma_{\ell n}}{2} = \frac{d-3}{\sqrt{d-1}}\sqrt{1-\frac{\theta^2}{r_c^{2(d-3)}}}
\sqrt{1-\frac{(d-2)\theta^2}{r_c^{2(d-3)}}}\nonumber \\ && \qquad \times
\left(\frac{1}{r_{c}}\right)\alpha(n)+\underset{\ell \to +\infty}{\cal O}\left(\frac{1}{\ell}\right).
\end{eqnarray}
\end{subequations}
It should be noted that for $\theta=0$, from (\ref{PR_RNd}) and
(\ref{QNMRNd}),  we recover the Schwarzschild-Tangherlini BH results
of Sec.~IV.A.4.

\subsection{The canonical acoustic black hole}

For the canonical acoustic BH (for more details see Sec.~8 of Ref.~\cite{Visser98}), the function $f(r)$ reads
\begin{equation}
f(r)=1-\left(\frac{r_0}{r}\right)^{4}.
\end{equation}
The sonic event horizon is located at $r_h=r_0$, the radius of the
unstable circular null geodesics is given by
\begin{subequations}
\begin{equation}
r_c=3^{1/4}r_0
\end{equation}
and we have for the corresponding $\eta_c$ parameter
\begin{equation}
\eta_c=2.
\end{equation}
\end{subequations}
Furthermore, we have
\begin{subequations}
\begin{eqnarray}
&&a_{n}=\frac{1}{108}+\frac{20}{9}\alpha(n)^{2},\\
&&b_{n}=-\frac{85}{1944}-\frac{215}{243}\alpha(n)^{2},\\
&&c_{n}=-\frac{19}{243}+\frac{110}{243}\alpha(n)^{2},\\
&&\epsilon_{n}(\omega)=\frac{b_{n}}{(3/2)r_{c}^{2} \, \omega^{2}+a_{n}+4\alpha(n)^{2}}.
\end{eqnarray}
\end{subequations}
The WKB approximation (\ref{PRapproxWKB}) for the Regge poles leads to
\begin{eqnarray}
&&\lambda_{n}(\omega) \approx
\left[\frac{3}{2}r_{c}^{2}\omega^{2}+a_{n}
+8\alpha(n)^{2}\epsilon_{n}(\omega)\right]^{1/2}\nonumber \\ &&
\qquad+2i\alpha(n)\left[1+\epsilon_{n}(\omega)\right]
\end{eqnarray}
and their high frequency behavior (\ref{RPHF}) provides
\begin{eqnarray}
&&\lambda_{n}(\omega)=\left[\frac{r_c}{\sqrt{2}}\omega+\frac{a_{n}}{\sqrt{2}\,r_{c}\omega}\right]\nonumber \\
&& \qquad+2i\alpha(n)\left[1+\frac{b_{n}}{(3/2)r_{c}^{2}\omega^{2}}\right]
+\underset{\omega \to +\infty}{\cal O}\left(
\frac{1}{\omega^3}\right).\nonumber \\
&&
\end{eqnarray}

The resonance excitation frequencies and the damping of the QNMs given by the general formulas (\ref{asympQNM})
reduce to
\begin{subequations}\label{QNMCABH}
\begin{eqnarray}
&&\omega_{\ell n}^{(0)}=\frac{\sqrt{2}}{r_{c}}\left[(\ell+1/2)-\frac{a_{n}}{2\ell}+\frac{a_{n}}{4\ell^{2}}+\underset{\ell \to +\infty}{\cal O}\left(
\frac{1}{\ell^3}\right)\right],\nonumber \\
&& \label{QNMCABHa}\\
&&\frac{\Gamma_{\ell n}}{2}=\frac{2\sqrt{2}}{r_{c}}\alpha(n)\left[1+\frac{c_{n}}{2\ell^{2}}+\underset{\ell \to +\infty}{\cal O}\left(
\frac{1}{\ell^3}\right)\right]\label{QNMCABHb}.
\end{eqnarray}
\end{subequations}
The leading-order terms of (\ref{QNMCABH}) are in agreement with
formula (53) of Ref.~\cite{BertiCardosoLemos2004}. In
Ref.~\cite{DolanOttewill_2009}, Dolan and Ottewill have obtained for
$n=0$ the expansions of $\omega_{\ell n}^{(0)}$ and $\Gamma_{\ell
n}/2$ up to order $1/(\ell+1/2)^{4}$. Our results (\ref{QNMCABH})
are consistent with their Eq.~(78).

\section{Conclusion}

As noted by Chandrasekhar in the mid-1970s, BH perturbation theory
can be formulated as a resonant scattering problem. As a
consequence, all the techniques developed in the framework of
scattering theory can be naturally introduced in the context of BH
physics. One of the central concepts of scattering theory is the
concept of a Regge pole which permits one to obtain semiclassical
interpretations of resonance phenomena. We have recently used it in
Refs.~\cite{DecaniniFJ_cam_bh} and \cite{DecaniniFolacci2010a} in
order to understand, from a new point of view, some aspects of the
resonant Schwarzschild BH. Our results have permitted us to
established more particularly, on a rigorous basis and in the
wave/field theory context, the appealing and intuitive
interpretation of Schwarzschild BH QNMs suggested by Goebel in 1972
\cite{Goebel}, i.e., that they could be interpreted in terms of
gravitational waves in spiral orbits close to the unstable circular
photon/graviton orbit at $r=3M$ which decay by radiating away
energy.

In the present paper, we have greatly extended the approach
initiated in Ref.~\cite{DecaniniFolacci2010a}. More precisely, we
have provided general analytical formulas for the Regge poles of the
$S$-matrix associated with a massless scalar field theory defined on
a static spherically symmetric BH of arbitrary dimension with a
photon sphere. This has been achieved by using third-order WKB
approximations to solve the associated radial wave equation and by
emphasizing more particularly the role of the photon sphere or, in
other words, of the unstable circular null geodesics on which a
massless particle can orbit the BH. But, it is important to note
that our results are only a first step to understand, in a
semiclassical framework, various aspects of static spherically
symmetric BH physics such as wave scattering, gravitational lensing,
Hawking radiation... Moreover, we think that it would now be very
interesting to introduce the CAM method in a less symmetric
situation, e.g., to study Kerr and Kerr-Newman BHs (see
Ref.~\cite{GlampedakisAndersson2003} for a first numerical step in
this direction) and to analyze the splitting of their complex
quasinormal frequencies or to understand, from a semiclassical point
of view, the superradiance phenomenon. We leave such a study for the
future.

\appendix

\section{From Regge poles to complex quasinormal frequencies in arbitrary dimensions}

In this appendix, we begin by extracting the resonant part of the
form factor associated with the $S$-matrix defined in Sec.~II. We
then interpret it in terms of ``surface waves", each one associated
with a Regge pole. This last result permits us to derive the
semiclassical formulas (\ref{sc1})-(\ref{sc4}) which are crucial to
construct, in Sec.~III, from the Regge trajectories, the spectrum of
the complex frequencies corresponding to the weakly damped QNMs.

\begin{figure}
\includegraphics[height=8cm,width=8cm]{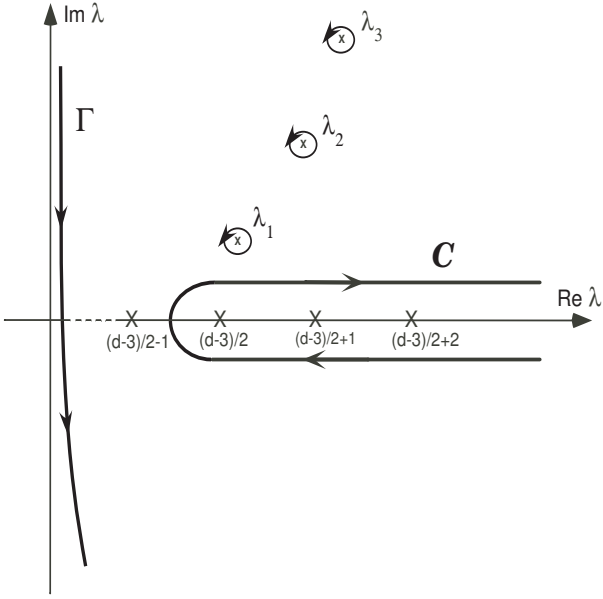}
\caption{\label{fig:watson} The Watson integration contour. }
\end{figure}

Let us consider the scattering of a monochromatic scalar plane wave
of frequency $\omega$ by a $d$-dimensional static and spherically
symmetric BH. Without loss of generality, the corresponding form
factor can be written in the form
\begin{eqnarray}\label{formfactor}
&& f(\omega, \theta_1)=\frac{2^{\frac{d-4}{2}}}{2\sqrt{\pi}(i
\omega)^{\frac{d-2}{2}}}
\Gamma\left(\frac{d-3}{2}\right)\nonumber \\
&& \quad \times \sum_{\ell=0}^{+\infty}(2\ell +
d-3)\left[S_\ell(\omega) - 1 \right]C_{\ell}^{\frac{d-3}{2}}(\cos
\theta_1)\nonumber \\
&&
\end{eqnarray}
where $\ell$ is the ordinary angular momentum index,
$C_{\ell}^{\frac{d-3}{2}}(z)$ are the Gegenbauer polynomials (see,
for example, Ref.~\cite{AS65}) and $S_\ell(\omega)$ are the diagonal
elements of the $S$-matrix defined by Eq.~(\ref{bc2}). By means of a
Sommerfeld-Watson transformation \cite{Watson18,Sommerfeld49} (see
for example Ref.~\cite{New82} for a more recent presentation), we
can extract in two steps the resonant part of the form factor
(\ref{formfactor}). We first replace the discrete sum over the
ordinary angular momentum $\ell$ by a contour integral in the
complex $\lambda$-plane. We obtain
\begin{eqnarray}\label{SWTransfo}
&&f(\omega, \theta_1)=i\, \frac{2^{\frac{d-4}{2}}}{2\sqrt{\pi}(i
\omega)^{\frac{d-2}{2}}}
\Gamma\left(\frac{d-3}{2}\right)\nonumber \\
&& \quad \times \int_{\cal C} \, \frac{\lambda\left[S_{\lambda -\frac{d-3}{2}
} (\omega) - 1 \right]}{\sin[ \pi(\lambda-(d-3)/2)]}C_{\lambda -\frac{d-3}{2}}^{\frac{d-3}{2}}
(-\cos \theta_1) ~d\lambda .\nonumber \\
&&
\end{eqnarray}
Here $\mathcal{C}$ is the integration contour in
the complex $\lambda$-plane displayed in Fig.~\ref{fig:watson}.
Furthermore, $S_{\lambda -(d-3)/2} (\omega)$ is now the analytic
extension of $S_\ell (\omega )$ into the complex $\lambda$-plane
(CAM plane) which is regular in the vicinity of the positive real
$\lambda$ axis, and $C_{\lambda -\frac{d-3}{2}}^{\frac{d-3}{2}}(z)$
is the analytical extension of the Gegenbauer polynomials which is
defined by \cite{AS65}
\begin{eqnarray}
&&C_{\lambda -\frac{d-3}{2}}^{\frac{d-3}{2}}(z)=\frac{\Gamma \left(\lambda
+\frac{d-3}{2}\right)}{{\Gamma
(d-3) \Gamma \left(\lambda -\frac{d-5}{2}\right)}}\nonumber \\
&& \times F\left(-
\lambda + \frac{d-3}{2},\lambda +
\frac{d-3}{2};\frac{d-2}{2};\frac{1-z}{2}\right).
\end{eqnarray}

We can then deform the path of integration in Eq.~(\ref{SWTransfo})
taking into account the possible singularities (see
Fig.~\ref{fig:watson}), i.e., the poles of the $S$-matrix lying in
the first quadrant of the CAM plane, or in other words, the Regge
poles $\lambda_n (\omega)$ with $n=1,2, \dots $ By Cauchy's theorem
we can then extract from Eq.~(\ref{SWTransfo}) a residue series over
Regge poles which permits us to obtain the resonant contribution in
the form
\begin{eqnarray}\label{ampliPR1a}
&&f_\mathrm{P}(\omega, \theta_1)=-2\pi\,\frac{2^{\frac{d-4}{2}}}{2\sqrt{\pi}(i \omega)^{\frac{d-2}{2}}}
\Gamma\left(\frac{d-3}{2}\right)\nonumber\\
&& \times \sum_{n=1}^{+\infty}\frac{\lambda_n(\omega
)r_n(\omega)}{\sin \left[\pi(\lambda_n(\omega )-(d-3)/2)\right]}C_{\lambda_n(\omega
)-\frac{d-3}{2}}^{\frac{d-3}{2}}(-\cos\theta_1).\nonumber \\
&&
\end{eqnarray}
Here we have defined
\begin{equation}
r_n(\omega )=\mathrm{residue}\left[S_{\lambda-(d-3)/2}(\omega
)\right]_{\lambda = \lambda _n(\omega )}.
\end{equation}
It should be noted that $f(\omega,\theta_1)$ differs from
$f_\mathrm{P}(\omega,\theta_1)$ by a background integral over the
contour $\Gamma$ (see Fig.~\ref{fig:watson}) which does not play any
role in resonance phenomena.

By using the asymptotic expansion (see Ref.~\cite{AS65})
\begin{eqnarray}
&&C_{\lambda-\frac{d-3}{2}}^{\frac{d-3}{2}}(-\cos \theta_1)
\sim \frac{\Gamma\left(\frac{d-2}{2}\right)\Gamma\left(\lambda
+\frac{d-3}{2}\right)}{\Gamma\left(d-3\right)\Gamma\left(\lambda+\frac{1}{2}\right)}\nonumber \\
&& \quad \times \frac{e^{i\lambda
(\pi-\theta_1)-i(d-3)\pi/4}+e^{-i\lambda
(\pi-\theta_1)+i(d-3)\pi/4}}{2[{\pi(\lambda-(d-3)/2)}]^{1/2}[({\sin
\theta_1})/2]^{(d-3)/2}}\nonumber \\
&&
\end{eqnarray}
as $|\lambda| \to \infty$, which is valid for $|\lambda| \sin
\theta_1
>1$, as well as the relation
\begin{eqnarray}
&&\frac{1}{\sin \pi z }=-2i \sum_{m=0}^{+\infty} e^{i\pi(2m+1)z}
\end{eqnarray}
which is true if $\mathrm{Im} \ z > 0$, we can write
\begin{widetext}
\begin{eqnarray}\label{ampliPR2}
&&f_\mathrm{P}(\omega, \theta_1)= \frac{2 i \pi}{(2i
\omega)^{\frac{d-2}{2}}}
\sum_{n=1}^{+\infty}\frac{\Gamma\left[\lambda_n(\omega)
+(d-3)/2\right]} {\Gamma\left[\lambda_n(\omega)+ 1/2
\right]}\frac{\lambda_n(\omega )
r_n(\omega)e^{-i\pi(d-3)/2}}{[{\pi(\lambda_n(\omega)-(d-3)/2)}]^{1/2}[({\sin
\theta_1})/2]^{(d-3)/2}}
\nonumber \\
&& \quad \times
\sum_{m=0}^{+\infty}\left[e^{i\lambda_n(\omega)(2\pi-\theta_1+2m\pi)
-i(d-3)m\pi-i(d-3)\pi/4}+e^{i\lambda_n(\omega)(\theta_1+2m\pi)-i(d-3)m\pi+i(d-3)\pi/4}\right].
\end{eqnarray}
\end{widetext}
In Eq.~(\ref{ampliPR2}), exponential terms correspond to ``surface
wave"/diffractive contributions. Keeping in mind the time dependence
$\exp(-i\omega t)$ and recalling that a given Regge pole
$\lambda_n(\omega)$ lies in the first quadrant of the CAM plane,
$\exp[i\lambda_n(\omega)(\theta_1)]$ (resp.~$\exp[i\lambda_n(\omega
)(2\pi-\theta_1)]$) corresponds to the so-called $n$th ``surface
wave" propagating counterclockwise (resp.~clockwise) around the BH
and $\mathrm{Re} \ \lambda_n(\omega)$ represents its azimuthal
propagation constant while $\mathrm{Im} \ \lambda_n(\omega)$ is its
damping constant. The corresponding exponential decay
$\exp[-\mathrm{Im} \ \lambda_n(\omega)\theta_1]$
(resp.~$\exp[-\mathrm{Im} \ \lambda_n(\omega)(2\pi-\theta_1)]$) is
due to continual reradiation of energy. Moreover, in
Eq.~(\ref{ampliPR2}), the sum over $m$ takes into account the
multiple circumnavigations of the ``surface waves" around the BH as
well as the associated radiation damping. Finally, it should be
noted that, in Eq.~(\ref{ampliPR2}), the presence of the factor
$\exp[-i(d-3)m\pi]$ accounts for the phase advance due to caustics.

The resonant behavior of the BH can now be understood in terms of
``surface waves". Let us consider the $n$th ``surface wave", i.e.,
the ``surface wave" described by the Regge pole $\lambda_n(\omega
)$. When the quantity $\mathrm{Re} \ \lambda_n (\omega ) - (d-3)/2$
coincides with an integer, a resonance occurs: it is produced by a
constructive interference between the different components of the
surface wave, each component corresponding to a different number of
circumnavigations of the BH [see Eq.~(\ref{ampliPR2})]. the
resonance excitation frequencies $\omega^{(0)}_{\ell n}$ appearing
in the Breit-Wigner formula (\ref{BW}) and generated by the $n$th
``surface wave" are therefore obtained from the
Bohr-Sommerfeld--type quantization condition
\begin{equation}\label{sc1_Appendix}
\mathrm{Re}  \, \lambda_n \left(\omega^{(0)}_{\ell n} \right)= \ell
+ \frac{d-3}{2},  \qquad \ell \in {\bf N}.
\end{equation}
Now, by assuming that $\omega $ is in the neighborhood of
$\omega^{(0)}_{\ell n}$, we can expand $\lambda_n (\omega)$ in a
Taylor series about $\omega^{(0)}_{\ell n}$ and write
\begin{eqnarray} \label{TS_PR}
& &  \lambda_n (\omega ) \approx \ell  + \frac{d-3}{2} +  \left.
\frac{d \, \mathrm{Re} \, \lambda_n (\omega )}{d\omega}
\right|_{\omega =\omega^{(0)}_{\ell n}} \left(\omega -
\omega^{(0)}_{\ell
n}  \right) \nonumber \\
& & \quad +  i \, \mathrm{Im} \, \lambda_n \left(\omega^{(0)}_{\ell
n}\right) + i \left. \frac{d \, \mathrm{Im} \,
\lambda_n(\omega)}{d\omega} \right|_{\omega =\omega^{(0)}_{\ell n}}
\left(\omega - \omega^{(0)}_{\ell n} \right) \nonumber
\\ &  &   \quad  + \dots
\end{eqnarray}
Then, by replacing (\ref{TS_PR}) in the term ${\sin \left[\pi(
\lambda_n(\omega )-(d-3)/2  )\right]}$ of (\ref{ampliPR1a}), we can
see that $f_\mathrm{P}(\omega, \theta_1)$ presents a resonant
behavior given by the Breit-Wigner formula (\ref{BW}) with
\begin{equation}\label{sc2_Appendix} \frac{\Gamma _{\ell n}}{2}= \left.  \frac{\mathrm{Im} \,
\lambda_n (\omega )[d/d\omega \, \mathrm{Re}\, \lambda_n (\omega )
]}{[d/d\omega \, \mathrm{Re} \, \lambda_n (\omega ) ]^2 + [d/d\omega
\, \mathrm{Im} \, \lambda_n (\omega ) ]^2 } \right|_{\omega
=\omega^{(0)}_{\ell n}}.
\end{equation}
Furthermore, it should be noted that in the frequency range where
the condition $| d/d\omega \,\mathrm{Re} \, \lambda_n (\omega )  |
\gg |d/d\omega \, \mathrm{Im}\, \lambda_n (\omega )  |$ is
satisfied, (\ref{sc2_Appendix}) reduces
to\begin{equation}\label{sc4_Appendix} \frac{\Gamma _{\ell n}}{2}=
\left.  \frac{\mathrm{Im} \, \lambda_n (\omega )}{d/d\omega \,
\mathrm{Re} \, \lambda_n (\omega )  } \right|_{\omega
=\omega^{(0)}_{\ell n}}.
\end{equation}

From the Regge trajectories, i.e., the curves traced out in the CAM
plane by the functions $\lambda_n (\omega )$ for $\omega \in
[0,+\infty[$, formulas (\ref{sc1_Appendix}) and (\ref{sc2_Appendix})
or (\ref{sc4_Appendix}) permit us to construct semiclassically the
resonance spectrum of the BH or, in other words, the spectrum of the
quasinormal complex frequencies. Equation (\ref{sc1_Appendix})
provides the location of the excitation frequencies
$\omega^{(0)}_{\ell n}$ of these complex resonances while
Eq.~(\ref{sc2_Appendix}) provides the corresponding widths. Of
course, the reasoning leading to (\ref{sc1_Appendix}) and
(\ref{sc2_Appendix}) from (\ref{formfactor}) is based on many
assumptions that are not necessarily satisfied in practice. But, in
general, it permits one to describe rather correctly quasinormal
complex frequencies lying near the real $\omega$-axis.

\section{Regge poles and QNMs of the Schwarzschild-de Sitter black
hole}
For the $d$-dimensional Schwarzschild-de Sitter BH \cite{Tangherlini63}, the function
$f(r)$ reads
\begin{equation} \label{fr_SdSBH}
f(r)=1-\left(\frac{r_0}{r}\right)^{d-3}-\frac{r^{2}}{L^{2}}.
\end{equation}
Here, $r_0$ is a parameter associated with the mass $M$ of the BH by
Eq.~(\ref{r0dimd}) and $L$ is a characteristic length linked to the
cosmological constant $\Lambda$ by the relation
\begin{equation}
\Lambda=\frac{(d-1)(d-2)}{2L^{2}}.
\end{equation}
If
\begin{eqnarray}\label{condhoriz}
&&0<\left(\frac{d-1}{2}\right)^{2/(d-1)}\left[\left(\frac{d-1}{d-3}\right)
\frac{r_0^2}{L^2}\right]^{(d-3)/(d-1)}<1\nonumber \\
&&
\end{eqnarray}
the equation $f(r)=0$ has only two real roots $r_h$ and
$r_\mathrm{Co}$ with $0<r_h<r_\mathrm{Co}$ (see, for example,
Appendix A of Ref.~\cite{NatarioSchiappa2004}). They correspond to
the radii of the BH horizon and of the cosmological horizon. For $r
\in ]r_h,r_\mathrm{Co}[$, we have $f(r)>0$. In that case, there also
exists a photon sphere located at
\begin{subequations}
\begin{equation}\label{rcSchTangDeSitter}
r_c=r_0\left(\frac{d-1}{2}\right)^{1/(d-3)}
\end{equation}
with $r_c \in ]r_h,r_\mathrm{Co}[$ and the associated $\eta_c$ parameter is given by
\begin{equation}\label{etaSchTangDesitter}
\eta_c=\sqrt{d-3}.
\end{equation}
\end{subequations}
$r_c$ and $\eta_c$ are both independent of the cosmological constant
and are those already obtained for the ordinary
Schwarzschild-Tangherlini BH [compare with Eqs.~(\ref{rcSchTang})
and (\ref{etaSchTang})].

Here, it is important to note that, at first sight, the formalism we
have previously developed does not apply to this gravitational
background which is not asymptotically flat and for which the
function $f(r)$ is positive only in the finite interval $]r_h,r_\mathrm{Co}[$.
In particular, the proof of the semiclassical formulas
(\ref{sc1})-(\ref{sc4}) given in Appendix A is not valid because
it is based on the notions of $S$-matrix and form factor which, here,
cannot be used. But we can easily circumvent these difficulties:
\begin{itemize}
\item Indeed, for $f(r)$ given by (\ref{fr_SdSBH}) the potential $V_\ell(r)$
obtained from (\ref{EffectivePot}) satisfies the boundary conditions
$\lim_{r \to r_h} V_\ell(r)=0$ and $\lim_{r \to r_\mathrm{Co}}
V_\ell(r)=0$. As the tortoise coordinate $r_\ast(r)$ provides a
bijection from $]r_h,r_\mathrm{Co}[$ to $]-\infty,+\infty[$, it is
formally possible to define a kind of $S$-matrix from the relations
(\ref{bc1}) and (\ref{bc2}).
\item The structure of this $S$-matrix in the complex $\omega$-plane and in the
complex $\lambda$-plane allows us to consider the spectra of the quasinormal frequencies
and of the Regge poles.
\item It is then possible to establish a semiclassical connection based on (\ref{sc1})-(\ref{sc4})
between these two spectra. This can be achieved by constructing,
from the Regge modes, the diffractive part of the Feynman propagator
associated with the scalar field, i.e., by extending a formalism
introduced a long time ago by Sommerfeld \cite{Sommerfeld49} as an
alternative to the usual approach of scattering \cite{Watson18}
considered in Appendix A. Such an approach has been used in
Ref.~\cite{DecaniniFolacci2009} for the BTZ BH.
\end{itemize}

\subsection{The four-dimensional Schwarzschild-de Sitter black hole}
For $d=4$, the condition (\ref{condhoriz}) implies
\begin{equation}
0<\frac{27M^2}{L^2}<1
\end{equation}
and we have
\begin{subequations}\label{rc_eta_SdS4}
\begin{eqnarray}
&&r_{c}=3M,\\
&&\eta_c=1,
\end{eqnarray}
\end{subequations}
as well as
\begin{subequations}
\begin{eqnarray}
&&a_{n}=-\frac{29}{216}+\frac{137M^{2}}{8L^{2}}+\frac{5}{18}\left(1-\frac{27M^{2}}{L^{2}}\right)
\alpha(n)^{2},\nonumber \\
&&\\
&&b_{n}=\frac{1357}{15552}+\frac{1505M^{2}}{288L^{2}}-\frac{24765M^{4}}{64L^{4}}\nonumber \\
&&\qquad+\left[-\frac{305}{3888}+\frac{35M^{2}}{72L^{2}}
+\frac{705M^{4}}{16L^{4}}\right]\alpha(n)^{2},\\
&&c_{n}=\frac{313}{7776}+\frac{3971M^{2}}{144L^{2}}-\frac{24765M^{4}}{32L^{4}}\nonumber \\
&&\qquad+\left[\frac{235}{1944}-\frac{235M^{2}}{36L^{2}}
+\frac{705M^{4}}{8L^{4}}\right]\alpha(n)^{2},\\
&&\epsilon_{n}(\omega)=\frac{b_{n}}{27M^{2}/(1-27M^{2}/L^{2})
\, \omega^{2}+a_{n}+\alpha(n)^{2}}.\nonumber \\
&&
\end{eqnarray}
\end{subequations}
The WKB approximation (\ref{PRapproxWKB}) for the Regge poles leads to
\begin{eqnarray}\label{PRWKB_SdS_4}
&&\lambda_{n}(\omega) \approx
\left[\frac{27M^{2}}{1-27M^{2}/L^{2}}\omega^{2}
+a_{n}+2\alpha(n)^{2}\epsilon_{n}(\omega)\right]^{1/2}\nonumber \\
&& \qquad +i\alpha(n)\left[1+\epsilon_{n}(\omega)\right]
\end{eqnarray}
and their high frequency behavior (\ref{RPHF}) provides
\begin{eqnarray}\label{PRWKB_HF_SdS_4}
&&\lambda_{n}(\omega)=\left[\frac{3\sqrt{3}M}{\sqrt{1-27M^{2}/L^{2}}}\omega+\frac{\sqrt{1-27M^{2}/L^{2}}\,a_{n}}{6\sqrt{3}M\omega}\right]\nonumber
\\
&&+i\alpha(n)\left[1+\frac{(1-27M^{2}/L^{2})b_{n}}{27M^{2}\omega^{2}}\right]+\underset{\omega
\to +\infty}{\cal O}\left( \frac{1}{\omega^3}\right).\nonumber \\
&&
\end{eqnarray}

Here, it is worth noting that, while the radius of the photon sphere
and the $\eta_c$ parameter characterizing the instability of the
circular null geodesics do not depend on the cosmological constant
[see formulas (\ref{rc_eta_SdS4})], this is not the case as regards
the properties (dispersion relation and damping) of the ``surface
waves" lying close to the photon sphere [see
Eqs.~(\ref{PRWKB_SdS_4}) and (\ref{PRWKB_HF_SdS_4})]. As a
consequence, the cosmological constant affects strong gravitational
lensing by the four-dimensional Schwarzschild-de Sitter BH. This
result corroborates, in a semiclassical framework, the recent
analysis of Rindler and Ishak \cite{RindlerIshak2007} concerning the
role that the cosmological constant plays in the bending of light
around a concentrated mass (for references on this hot subject, we
refer to the bibliography of the recent paper by Ishak, Rindler and
Dossett \cite{IshakRindlerDossett2010}).

The resonance excitation frequencies and the damping of the QNMs given by the
general formulas (\ref{asympQNM}) reduce to
\begin{subequations}\label{QNMSchDeSitter4}
\begin{eqnarray}
&&\omega_{\ell n}^{(0)}=\frac{\sqrt{1-27M^{2}/L^{2}}}{3\sqrt{3}M}\nonumber \\
&& \quad \times \left[\left(\ell+\frac{1}{2}
\right)-\frac{a_{n}}{2\ell}+\frac{a_{n}}{4\ell^2}+\underset{\ell \to
+\infty}{\cal O}\left(
\frac{1}{\ell^3}\right)\right],\\
&&\frac{\Gamma_{\ell n}}{2}=\frac{\sqrt{1-27M^{2}/L^{2}}}{3\sqrt{3}M}\alpha(n)
\left[1+\frac{c_{n}}{2\ell^{2}}+\underset{\ell
\to +\infty}{\cal O}\left(
\frac{1}{\ell^3}\right)\right].\nonumber \\
&&
\end{eqnarray}
\end{subequations}
The leading-order terms of (\ref{QNMSchDeSitter4}) have been
obtained in Ref.~\cite{BarretoZworski1997}. In
Ref.~\cite{DolanOttewill_2009}, Dolan and Ottewill have obtained for
$n=0$ the expansions of $\omega_{\ell n}^{(0)}$ and $\Gamma_{\ell
n}/2$ up to order $1/(\ell+1/2)^{4}$. Our results
(\ref{QNMSchDeSitter4}) are consistent with their Eq.~(72). Finally,
it should be noted that, for $L \to +\infty$, all the results
obtained for the Regge poles and the complex quasinormal frequencies
of the four-dimensional Schwarzschild-de Sitter BH reduce to the
Schwarzschild BH results of Sec.~IV.A.1.

\subsection{Leading-order terms for the Schwarzschild-de Sitter black hole of arbitrary dimension}

In the $d$-dimensional case, we can easily derive the leading-order
terms of the Regge poles and quasinormal complex frequencies. We
have
\begin{eqnarray}\label{RP_SchDeSitter}
&&\lambda_{n}(\omega)=\sqrt{\frac{d-1}{d-3}}\frac{r_{c}\omega}{\sqrt{1-\frac{(d-1)r_c^2}{(d-3)L^2}}}
\nonumber \\ && \qquad  +i\sqrt{d-3}\,\alpha(n)+\underset{\omega \to
+\infty}{\cal O}\left(\frac{1}{\omega}\right)
\end{eqnarray}
and
\begin{subequations}\label{QNMSchDeSitter}
\begin{eqnarray}
&&\omega_{\ell n}^{(0)} = \sqrt{\frac{d-3}{d-1}}\sqrt{1-\frac{(d-1)r_c^2}{(d-3)L^2}}
\left(\frac{1}{r_{c}}\right)\left(\ell+\frac{d-3}{2}\right)\nonumber \\
&& \qquad +\underset{\ell \to +\infty}{\cal O}\left(
\frac{1}{\ell}\right),\\
&&\frac{\Gamma_{\ell n}}{2} = \frac{d-3}{\sqrt{d-1}}\sqrt{1-\frac{(d-1)r_c^2}{(d-3)L^2}}
\left(\frac{1}{r_{c}}\right)\alpha(n)\nonumber \\ && \qquad +\underset{\ell \to
+\infty}{\cal O}\left(\frac{1}{\ell}\right).
\end{eqnarray}
\end{subequations}
Formulas (\ref{QNMSchDeSitter}) are in agreement with the results
obtained in Ref.~\cite{ZerbiniVanzo2004}. Furthermore, it should be
noted that for $L \to +\infty$, from (\ref{RP_SchDeSitter}) and
(\ref{QNMSchDeSitter}), we recover the Schwarzschild-Tangherlini BH
results of Sec.~IV.A.4.

\bibliography{RP_SSBH}

\end{document}